\documentclass[fleqn,usenatbib]{mnras}
\usepackage{newtxtext,newtxmath}
\usepackage[T1]{fontenc}
\usepackage{threeparttable}
\DeclareRobustCommand{\VAN}[3]{#2}
\let\VANthebibliography\thebibliography
\def\thebibliography{\DeclareRobustCommand{\VAN}[3]{##3}\VANthebibliography}

\usepackage{graphicx}	% Including figure files
\usepackage{amsmath}	% Advanced maths commands
\usepackage{amssymb}	% Extra maths symbols

\usepackage{caption}
\title[Cold Gas with the Wind]{Not gone with the Wind: Survival of High-Velocity Molecular Clouds in the Galactic center}

\author[Mengfei Zhang, et al.]{
Mengfei Zhang,$^{1}$\thanks{E-mail: murphychang@zju.edu.cn}
Miao Li,$^{1}$\thanks{E-mail: miaoli@zju.edu.cn}
\\
$^{1}$School of Physics, Zhejiang University, Hangzhou, Zhejiang 210023, China\\
}

\begin{document}

\label{firstpage}
\pagerange{\pageref{firstpage}--\pageref{lastpage}}
\maketitle

\begin{abstract}
High-velocity atomic clouds in the Galactic center have attracted significant attention due to their enigmatic formation process, which is potentially linked to the starburst or supermassive black hole activities in the region. Further, the discovery of high-velocity molecular clouds (HVMCs) presents a greater puzzle, because they are much denser and more massive. If the HVMCs were accelerated by the strong activities in the Galactic center, they are expected to be destroyed before they reach such a high velocity. To shed light on this phenomenon, we perform three-dimensional numerical simulations to investigate the origin and hydrodynamic evolution of HVMCs during a starburst in the Galactic center. We find that the presence of a magnetic field provides effective protection and acceleration to molecular clouds (MCs) within the galactic winds. Consequently, the MCs can attain latitudes of approximately 1 kpc with velocities around 200 km s$^{-1}$, consistent with the observed characteristics of HVMCs. The consistency of our findings across a wide parameter space supports the conclusion that HVMCs can indeed withstand the starburst environment in the Galactic center, providing valuable insights into their survival mechanisms.

\end{abstract}


\section{Introduction}\label{sec:intro}

Galactic feedback, especially the nuclear wind, is now commonly accepted as an important process affecting the galactic evolution \citep[e.g.][and references therein]{2012ARA&A..50..455F, 2014ARA&A..52..589H, 2017hsn..book.2431H, 2017ARA&A..55...59N, 2018Galax...6..114Z}, which is, however, pretty weak in our Milky Way at present \citep{2003ApJ...591..891B, 2016A&A...589A..66H}.
Therefore, it is expected that Milky Way had been active before, but quenched after that, which should produce some corresponding relics.
Over the past tens of years, these feedback relics possibly have been discovered at radio, X-ray and $\gamma$-ray band, such as the Galactic Center Lobe (GCL; \citealp{1984Natur.310..568S}), the microwave haze \citep{2004ApJ...614..186F, 2013A&A...554A.139P}, the polarized lobes \citep{2013Natur.493...66C}, the Fermi bubbles \citep{2010ApJ...724.1044S}, the radio bubbles \citep{2019Natur.573..235H}, the X-ray chimneys \citep{2019Natur.567..347P} and the eROSITA bubbles \citep{Predehl2020}.
These structures have scales ranging from $\sim$100 pc to $\sim$10 kpc, indicating that they originated from a series of violent activities.
In addition, in the Galactic center, many high-velocity clouds (HVCs) were detected both above and below the Galactic plane \citep{2004ApJ...605..216C, 2005ApJ...623..196C, 2018ApJ...855...33D, 2020ApJ...888...51L, 2020ApJ...898..128A}. Especially, two high-velocity molecular clouds (HVMCs) are also discovered insides the HVCs \citep{2020Natur.584..364D},
The altitudes of the two MCs are 0.6 and 0.9 kpc, respectively.
Their velocities along $z$-axis are $\sim$ 180 and 150 km s${-1}$, while the radial velocities are $\sim$ 240 and 300 km s$^{-1}$ based on a biconical model \citep{2020Natur.584..364D}.
Their molecular mass are both $\sim$ 380 M$_{\odot}$, and their atomic mass are 220 and 800 M$_{\odot}$.
The HVMCs show good coincidence with some aforementioned relics, so they possibly originate from similar process, e.g., accelerated by the Galactic nuclear wind.
%Comparing with extra galaxies, Milky Way provides us a unique opportunity to simultaneously study the starburst, the AGN activity, the HVCs and the feedback relics in detail.

Although these relics and HVCs/HVMCs are commonly suggested to be produced by the feedback activity, the detailed mechanism is still the subject of intense debate.
Several models have been proposed to explain their formation, some of which focus on one structure \citep{2011PhRvL.106j1102C, Zubovas2011, Guo2012, 2012MNRAS.424..666Z, Fujita2013, Mou2014, Fujita2014, 2014MNRAS.444L..39L, Mou2015, Sarkar2015, Zhang2020}, while others attempt to simultaneously explain multiple structures \citep{2013MNRAS.436.2734Y, Crocker2015, Yang2017, 2021ApJ...913...68Z, 2022NatAs...6..584Y}.
Most of these models exhibit self-consistency, and some following simulations have provided further validation of their viability \citep[e.g.]{Guo2012, Mou2014, Mou2015, Sarkar2015, Zhang2020, 2022NatAs...6..584Y}.
However, simulating the acceleration of HVMCs  still presents a challenge for their formation models.
Compared to atomic clouds, molecular clouds are denser and cooler, making it more difficult to accelerate them to high velocity without disruption \citep{2017ApJ...834..144S, 2021ApJ...923L..11C}.
Some simulations for extra-galactic interaction between clouds and nuclear winds show  that clouds can be  protected by magnetic field \citep{2014MNRAS.444..971A, 2015MNRAS.449....2M, 2016MNRAS.455.1309B, 2017MNRAS.468.4801Z, 2020MNRAS.499.4261S, 2023MNRAS.tmp.1171J}, cooling \citep{2018MNRAS.480L.111G, 2020MNRAS.492.1970G, 2021MNRAS.501.1143K} and thermal conduction \citep{2017MNRAS.470..114A}, which confirms that cool clouds can survive acceleration by a hot wind.
Nevertheless, these simulations usually involve a constant hot wind, which is completely different from the unpredictable nuclear wind produced by starburst or AGN.
Moreover, most of them focus on high-latitude atomic or even ionized clouds, so they cannot clearly explain the formation of HVMCs at $\sim$ 1 kpc in our Milky Way.
It is therefore necessary to perform robust simulations to see whether the HVMCs observed in the Galactic center can be reproduced.

The formation of HVMCs is closely linked to the other feedback relics, and could potentially be used to distinguish among different models for their origin.
While the activity of active galactic nuclei (AGN) has the capability to accelerate molecular clouds (MCs) to high velocities, it is often so powerful that the clouds usually diffuse to atomic/ionized form. %it is often too powerful to sustain the existence of these clouds. (it is often too powerful to make the clouds diffuse to atomic/ionized form)
Unless, there are some periodic bursts, such as those arising from accretion onto the supermassive black hole (SMBH) Sgr A* \citep{2013Sci...341..981W}.
These bursts should be weaker than normal AGN, but still release comparable amounts of energy, allowing the MCs to be efficiently accelerated without being quickly destroyed.
% In addition, the atomic clouds are distributed over 20$\degree$ ($\sim$3 kpc) at longitude, while the longitude of HVMCs is 357 $\sim$ 358$\degree$ \citep{2018ApJ...855...33D,2020Natur.584..364D}.
% It is hard for AGN jet to directly scatter clouds over such a large extent, but AGN wind can possibly do that.
Relatively speaking, a starburst is a more feasible explanation for the formation of HVMCs, as the Galactic center exhibited a higher star formation rate about 30 million years ago \citep{NoguerasLara2019} and the molecular outflow is universal in active star-forming galaxies \citep{2018Sci...361.1016S, 2020MNRAS.493.3081R, 2020ApJ...905...86S, 2021A&A...653A.172S, 2023ApJ...944..134B}.
In fact, although the supernova feedback is important in galaxy formation  \citep{2015ApJ...802...99K, 2016MNRAS.459.2311M, 2017ApJ...841..101L, 2019MNRAS.483.3363H}, its working mechanism has not been fully understood, which leads to a difficulty to understand the role of HVMCs and also limits the cosmological simulations \citep{2020ApJ...890L..30L}.
%Supernova feedback is commonly studied on a galactic scale to draw general conclusions.
Currently, it is known that randomly distributed SNe in the disk only drive inefficient galactic winds because most supernova remnants lose their energy radiatively before breaking out of the disc \citep{2018MNRAS.481.3325F}, leading to a difficulty to push the HVMCs to high latitude in such a galactic wind.
Nevertheless, a starburst in the Galactic center can produce much stronger galactic wind and more efficiently accelerate the HVMCs.
It is expected that the starburst ended recently, but left these feedback relics and HVMCs.
It is difficult to tell which model is correct, because the hydrodynamical evolution of  AGN and starburst activity can be similar at large scale.
Their energy input rate can be similar, as a result, the wind driven by these activities can reach a comparable velocity at high latitude.
However, there should be noticeable differences at smaller scale ($\leq$ 1 kpc), such as the acceleration process of HVMCs and the morphology of relics, because the starburst can happen more randomly in a much larger region than the AGN activity.
Moreover, the metallicity of HVMCs is possibly different for AGN and starburst models, since starburst can produce more heavy elements.
Although \citet{2022NatAs...6..968A} indeed found different metallicitiy distribution of HVCs in the Galactic center, they explain that HVCs originate in Milky Way's disk and halo.
These models can be further examined through simulations.

In this paper, we investigate whether HVMCs observed in our Milky Way can be accelerated to high latitudes by a starburst. To this end, we perform a detailed simulation of the process.
We start by simulating a series of random core-collapse supernova explosions in the Galactic center with a frequency estimated based on a past star formation rate \citep{NoguerasLara2019}. We set a molecular cloud above the explosion region to study how the cloud is accelerated by the outflow wind and whether it can survive until it reaches 1 kpc, a position similar to the clouds detected by \citet{2020Natur.584..364D}. The explosion region is believed to be adjacent to the central molecular zone (CMZ), where more giant molecular clouds steadily exist.
Next, we check the density, temperature, and velocity of the clouds obtained from the simulations and modify initial conditions to study the influence of different parameters.
We will try to identify various HVMCs candidates obtained from the simulation by comparing with the observation and study their properties in detail.
Finally, we investigate the mixture of clouds and the ejecta of supernovae to disentangle the metallicity in HVMCs.

This paper will describe the simulation setup in Section~\ref{sec:sim} and show the results in Section~\ref{sec:res}.
The formation of the HVMCs, their metallicity and their relation with feedback relics will be discussed in Section~\ref{sec:dis}.
The Section~\ref{sec:sum} is a summary.

\section{Simulation}\label{sec:sim}

To perform the simulations, we utilized the publicly available, modular magnetohydrodynamic (MHD) code \textit{PLUTO}\footnote{http://plutocode.ph.unito.it/} \citep{Mignone2007, Mignone2012} to perform the simulations.
This grid-based MHD code employs a second-order Runge–Kutta time integrator and a Harten-Lax-van Leer Riemann solver for middle contact discontinuities, making it well-suited for simulating the interaction between the SN shock and the molecular clouds.

\subsection{Basic configuration}\label{subsec:config}
The simulation is based on a three-dimensional (3D) MHD cartesian frame with a grid of $\rm 200 \times 200 \times 2000$, equivalent to a physical volume of $\rm 100 \times 100 \times 1000$~pc$^3$ and a linear resolution of 0.5 pc pixel$^{-1}$.
% To cover a large simulation box, we have to use a relatively low resolution, which will possibly lead to an artificial protection for the clouds \citep{2017ApJ...834..144S}, but this will not influence the conclusion and we will discuss this in Section~\ref{subsec:form}.
We set the $z$-axis to be perpendicular to the Galactic disk (north as positive), the $y$-axis to run along decreasing Galactic longitude, and the $x$-axis to be parallel to the line-of-sight (the observer at the negative side).
We adopted an outflow boundary condition for all directions, which means that some of the clouds' material may flow outside of the simulation box.

The simulation is governed by the ideal MHD conservation equations,
\begin{eqnarray}
      \begin{cases}
      \dfrac{\partial \rho}{\partial t} + \nabla \cdot (\rho \bf{v}) = 0 ,\\
      \dfrac{\partial (\rho\bf{v})}{\partial t}+\nabla \cdot\left[\rho\bf{vv}+\bf{1}p\right]^{T}=-\rho \nabla \Phi, \\
      \dfrac{\partial E_{t}}{\partial t}+\nabla \cdot\left[\left(\dfrac{\rho \bf{v}^{2}}{2}+\rho \epsilon+p+\rho \Phi\right) \bf{v}- \dfrac{\bf{v} \times \bf{B} \times \bf{B}}{4\pi}\right] \\
      = -\dfrac{\partial\left( \rho \Phi\right)}{\partial t}, \\
      \dfrac{\partial \bf{B}}{\partial t} - \nabla \times (\bf{v} \times \bf{B}) = 0,
      \end{cases}
\end{eqnarray}
where $\rho$ is the mass density, $p$ the thermal pressure, $\bf{v}$ the velocity, $\bf{B}$ the magnetic field, $\bf{1}$ the dyadic tensor, $\Phi$ the gravitational potential, and $E_t$ the total energy density, defined as:
\begin{eqnarray}
  E_t = \rho \epsilon + \frac{(\rho\bf{v})^2}{2\rho} + \frac{\bf{B}^2}{8\pi},
\end{eqnarray}
where $\epsilon$ is the internal energy.
We use an ideal equation of state, i.e., $\epsilon = p/ (\Gamma -1)$, in which the ratio of specific heats $\Gamma$ = 5/3.

To accurately model the gravitational potential in the simulation volume, we assume that it is static and fully determined by the SMBH, the nuclear star cluster (NSC), and the nuclear disk (ND).
A point mass of $4\times10^6\rm~M_{\odot}$ is taken to represent the SMBH.
For the NSC and the ND, we adopt a spherical distribution following
\citet[Equation 5 therein]{2015MNRAS.447..948C}.
To incorporate radiative cooling in the simulation, we use a piece-wise cooling function with a lower limit of the cooling temperature set to 100 K.
We assume a solar abundance (H abundance $X_{\odot}$=0.711, He abundance $Y_{\odot}$=0.2741, metallicity $Z_{\odot}$=0.0149) for the ISM and the initial MC.
The multiphase gas in the Galactic center includes hot ionized ($\rm \sim 10^6$ K) \citep{2013ApJ...779...57K,2019Natur.567..347P}, warm ionized (10$^4$ to 10$^5$ K) \citep{2015ApJ...799L...7F, Bordoloi2017} and cool atomic (10$^3$ to 10$^4$ K) gas\citep{2013ApJ...770L...4M, 2018ApJ...855...33D}, etc., in which the gas lower than 100 K is usually taken as molecular gas.
In the simulation, temperatures below 100 K are typically not due to cooling, but rather due to adiabatic expansion.

\subsection{Supernova explosion and molecular clouds}\label{subsec:sw}
The initial conditions for our simulations are based on both observations and analytical models. Observationally, the high-velocity molecular clouds HVMCs typically exhibit densities ranging from 10 to 300 cm$^{-3}$ and outflow velocities between 200 and 300 $\rm km~s^{-1}$ \citep{2020Natur.584..364D}. However, to account for the significant gas loss that occurs during their propagation, we assume that the initial densities of the MCs should be higher. In addition, we need to consider other parameters such as the supernova explosion frequency and the initial latitude of the MCs to ensure that they reach the observed velocities without being completely destroyed. Therefore, we perform a systematic exploration of the parameter space to identify the most plausible initial conditions for our simulations.
Here, we introduce the cloud crushing time,
\begin{eqnarray}
      t_{\rm cc} = \dfrac{r_{\rm mc}}{v_{\rm sn}}\sqrt{\dfrac{\rho_{\rm mc}}{\rho_{\rm sn}}},
\label{eqn:tcc}
\end{eqnarray}
to quantify the timescale of cloud crushing \citep{1994ApJ...420..213K}, in which $r_{\rm mc}$ is the radius of the initial molecular cloud, $\rho_{\rm mc}$ the density of the cloud, $v_{\rm sn }$ the wind velocity produced by supernovae, and $\rho_{\rm sn}$ the wind density.
Based on general understanding, a cloud should begin to crush when the evolution time is longer than $t_{\rm cc}$, and should totally crush after a period of 2$t_{\rm cc}$. However, this estimation does not take into account the effects of the magnetic field and cooling mechanisms, which can play an important role in the cloud's evolution.

In a cylindrical region with a radius of 35 pc and a height of 10 pc, the fiducial SN birth rate is set to be $10\rm~kyr^{-1}$ \citep{2018ApJ...855...33D}, which is estimated by assuming an SFR of 1 M$_{\odot}$ yr$^{-1}$, a \citet{Kroupa2001} initial mass function (IMF) and a minimum mass of 8 M$_{\odot}$ for the progenitor star of a core-collapse SN.
The center of the cylindrical region is set to be located at the western 100 pc of Sgr A*.
\citet{Barnes2017} and \citet{2020MNRAS.497.5024S} estimated a current SFR of 0.1 M$_{\odot}$ yr$^{-1}$ inside the CMZ, while \citet{NoguerasLara2019} found that star formation in the ND (which has a similar radial extent as the CMZ) has been relatively active in the past 30 Myr, with an SFR of $0.2-0.8\rm~M_{\odot}~yr^{-1}$.
Our assumed SFR of 1 M$_{\odot}$ yr$^{-1}$ is compatible with a local starburst, which may be the case if SN events have been episodic and clustered on a $\lesssim$ Myr timescale.
This SFR is actually larger than the typical value in such a small region, so we also test a run with lower SN birth rate of $5\rm~kyr^{-1}$.
We have neglected Type Ia SNe, which have a birth rate of $\lesssim0.05\rm~kyr^{-1}$ according to the enclosed stellar mass in the ND/NSC \citep{2005A&A...433..807M}.
The SNe are set to randomly explode in the cylindrical region, and we use same random seed in all runs.

The density of the MCs follows an inverse square law, $\rm n_{mc} = n_0/r^2$, in which n$_0$ is the central density, r the radius.
% We also test a density profile of $\rm n_{mc} = n_{ISM} + (n_0-n_{ISM})/(1+(r/5)^{10})$ \citep{BandaBarragan2016}, which can naturally decrease to ISM density at the contact surface.
In the fiducial simulation, n$_0$ = 1500 H cm$^{-3}$, the maximum radius of the initial MC is 10 pc, and the height of the MC from the Galactic plane is 50 pc.
Based on these settings (r$_{\rm mc}$ = 10 pc, v$_{\rm sn}$ = 1000 km s$^{-1}$, n$_{\rm mc}$ = 15$\sim$50 H cm$^{-3}$, n$_{\rm sn}$ = 0.01 H cm$^{-3}$), we can estimate the $t_{\rm cc}$ 1$\sim$2 Myr, so the cloud will totally crush after 4 Myr in the classical analysis.
However, in our preliminary tests, we find the cloud can survive beyond 7 Myr by including a vertical magnetic field and the cooling effect. In this scenario, after around 7 Myr, the cloud will run outside of the simulation box.
Therefore, the simulation results are presented up until around 7 Myr.

In addition, once injected, the ejecta will eventually partially mix with the molecular clouds, and change their metallicity.
To study the mixture, we introduce two tracer parameters, $Q_1$ and $Q_2$, which are both evaluated at each pixel in the simulation and obey a simple conservation law:
\begin{eqnarray}
      \frac{\partial (\rho Q_i)}{\partial t} + \nabla \cdot (\rho Q_i \bf{v}) = 0.
\label{eqn:tracer}
\end{eqnarray}
$Q_1$ has a value of 1 for pure SN ejecta and 0 for the unpolluted molecular clouds and ISM, while
$Q_2$ has a value of 1 for pure molecular clouds and 0 for the unpolluted SN ejecta and ISM.
The values in between indicate a mixed gas.
These tracer parameters allow us to track the mixing process over time and analyze the distribution of metals in the simulated system.

\begin{table*}
\center
\caption{Parameters of the simulation runs}
\label{table:para}
\begin{threeparttable}
\begin{tabular}{cccccccccc} % four columns, alignment for each
\hline
Run & $I_{\rm SN}$ & $n_{\rm ism}$ & $n_{\rm mc}$ & $r_{\rm mc}$ & $h_{\rm mc}$ & $r_{\rm reg}$ & $h_{\rm reg}$ & $B$ \\
\hline
(1) & (2) & (3) & (4) & (5) & (6) & (7) & (8) & (9) \\
\hline
\textit{f100n1500v}   & 100  & 0.01  & 1500 & 10 & 50 & 35 & 10 & vertical magnetic field \\
\textit{f100n1500h}   & 100  & 0.01  & 1500 & 10 & 50 & 35 & 10 & horizontal magnetic field \\
\textit{f100n1500n}   & 100  & 0.01  & 1500 & 10 & 50 & 35 & 10 & no magnetic field \\
\textit{f100n1000v}   & 100  & 0.01  & 1000 & 10 & 50 & 35 & 10 & vertical  magnetic field \\
\textit{f200n1000v}   & 200  & 0.01  & 1000 & 10 & 50 & 35 & 10 & vertical  magnetic field \\
\hline
\end{tabular}
\begin{tablenotes}
\footnotesize
\item (1) Simulation run. (2) Explosion interval, in units of yr. (3) ISM H density, in units of $\rm cm^{-3}$. (4) The MC central H density, in units of $\rm cm^{-3}$. (5) The MC radius, in units of pc. (6) The MC height, in units of pc. (7) The radius of the cylindrical explosion region, in units of pc. (8) The height of the cylindrical explosion region, in units of pc.   (9) The direction of the magnetic field along the Galactic plane.
\end{tablenotes}
\end{threeparttable}
\end{table*}

\subsection{The ISM and the magnetic field}\label{subsec:runs}
We initialize our simulation with a uniform distribution of ISM density and temperature, with values of 0.01 H cm$^{-3}$ and 10$^6$ K, respectively, over the entire simulation box. Although thermal pressure is expected to be higher at lower latitudes due to rough hydrostatic equilibrium against gravity, our preliminary tests suggest that this effect is unimportant since the shock wave from the supernovae breaks this equilibrium early on.
Moreover, the stellar wind in the Galactic center is also strong and can unremittingly break this equilibrium.

The distribution of magnetic fields in the Galactic center remains a challenging problem, particularly in the central tens of parsecs \citep{Ferriere2009}, with many different components, influencing the strength and direction of the magnetic field.
There is actually a general model for the whole Milky Way \citep{2013lsmf.book..215B, 2017JCAP...10..019C}, in which the magnetic field is parallel to the Galactic plane at lower latitude and gradually tend to be perpendicular at higher latitude, but this is only an approximation in the Galactic center.
Therefore, we in this work test different runs, respectively with parallel, perpendicular and no magnetic field.

The magnetic strength range from $\sim$ 1 mG in the central tens of parsecs \citep{Ferriere2009} to few $\mu$G at 1 kpc above the Galactic plane \citep{2017JCAP...10..019C}.
For simplicity, we adopt a homogeneous magnetic strength of 10 $\mu$G over the whole simulation box.
The initial parameters are summarized in Table~\ref{table:para}.

\section{Results}\label{sec:res}

In this section, we present the simulation results.
We first describe in detail the evolution of the MCs in the vertical magnetic field in the fiducial run (Section~\ref{subsec:vB}) .
We then examine the role of the magnetic field in the two additional runs, one with horizontal magnetic field (Section~\ref{subsec:hB}) and the other with no magnetic field (Section~\ref{subsec:nB}), to illustrate how the change affects the formation of the HVMCs.
Finally, we study the influence of the cloud density and the supernovae explosion frequency (Section~\ref{subsec:rh}).

To quantitatively compare with the observation, we here parameterize the main features of the observed MCs, MW-C1 and MW-C2 \citep{2020Natur.584..364D} .
The altitudes of the two MCs are 0.6 and 0.9 kpc, respectively, so we choose $\sim$ 1 kpc as the standard position to guarantee the simulated clouds can indeed reach the height.
Their velocities along $z$-axis are $\sim$ 180 and 150 km s${-1}$, while the radial velocities are $\sim$ 240 and 300 km s$^{-1}$ based on a biconical model \citep{2020Natur.584..364D}.
Our simulation focuses on the propagation vertical to the Galactic plane, so we take 200$^{+100}_{-50}$ km s$^{-1}$ as the typical value.
The molecular mass of MW-C1 and MW-C2 are both $\sim$ 380 M$_{\odot}$, but their atomic mass are 220 and 800 M$_{\odot}$.
Thus we pay more attention to match the molecular mass, and the atomic mass can vary in a large range.
With a diameter of $\sim$30 pc, their mean molecular number densities are 130 and 190 H$_2$ cm$^{-3}$, and the mean atomic number densities are 1 and 3 H cm$^{-3}$.
In the work, we take the clouds denser than 10 H cm$^{-3}$ as MCs, and the clouds with a density between 1 and 10 H cm$^{-3}$  as atomic clouds.

In addition, there are also some qualitative features which are worth reproducing.
Surrounding the HVMCs, there are always some atomic clouds with lower density and larger volume, which were usually taken as HVCs before the discovery of HVMCs.
The number of detected HVCs is much larger than HVMCs, and most of HVCs are uniformly distributed above 250 pc \citep{2018ApJ...855...33D}.
There are possibly more HVMCs hidden in the HVCs, so more high-resolution and high-sensitivity molecular observations are necessary.

\subsection{The run for the fiducial set}\label{subsec:vB}

\begin{figure*}
\includegraphics[width=\textwidth,height=\textheight,keepaspectratio]{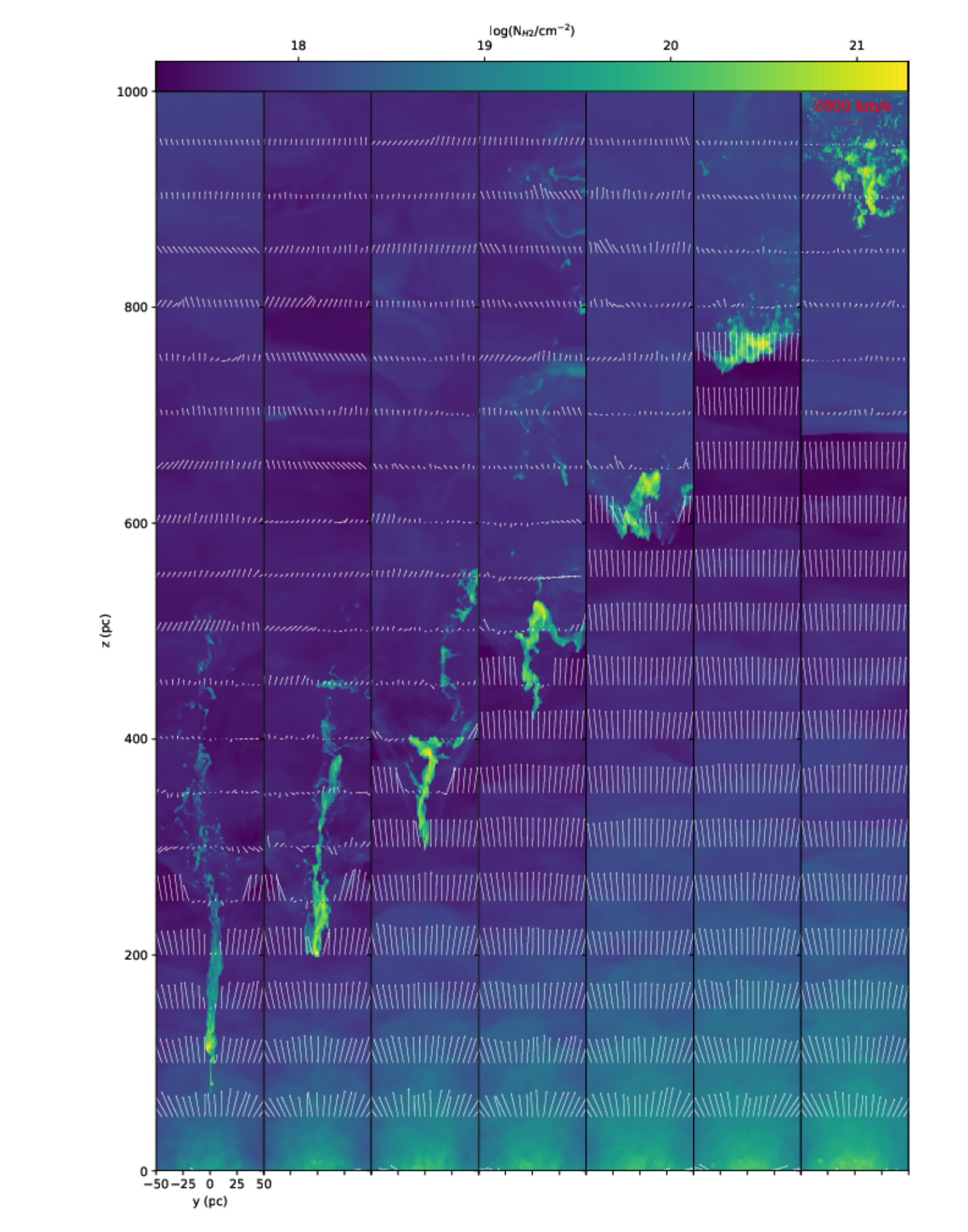}
\vskip-0.5cm
\caption{The $y$-$z$ column density maps of \textit{f100n1500v} between 1$\sim$7 Myr with a step of 1 Myr. The white arrows show the flow velocity in the slice through the $x$ = 0 pc, and the scale is shown at the upper right. The main cloud can indeed survive with comparable mass with the observation until it reaches 1 kpc at 7 Myr, though it will lose a large amount mass.
\label{fig:vB}}
\end{figure*}

\begin{figure*}
\includegraphics[width=\textwidth,height=\textheight,keepaspectratio]{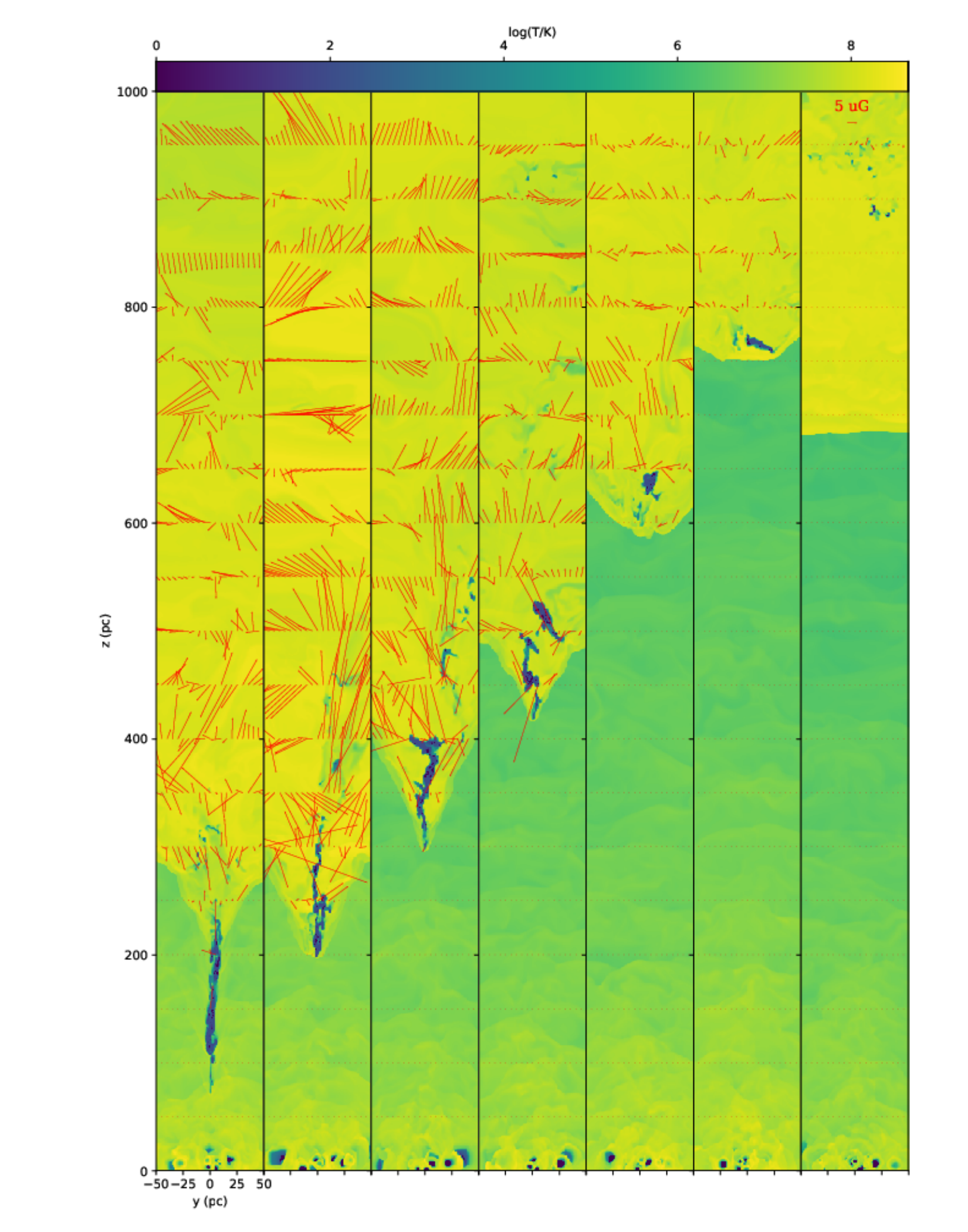}
\vskip-0.5cm
\caption{The temperature and the magnetic field maps of \textit{f100n1500v} in the slice through the $x$ = 0 pc between 1$\sim$7 Myr with a step of 1 Myr. The red arrows show the magnetic field, and the scale is shown at top right. The cold gas slowly diffuse away from the central slice, and almost dissipates at 7 Myr. The magnetic strength is amplified at the early stage, but gradually decreases after $\sim$ 3 Myr.
\label{fig:TvB}}
\end{figure*}

\begin{figure}
\includegraphics[width=0.5\textwidth]{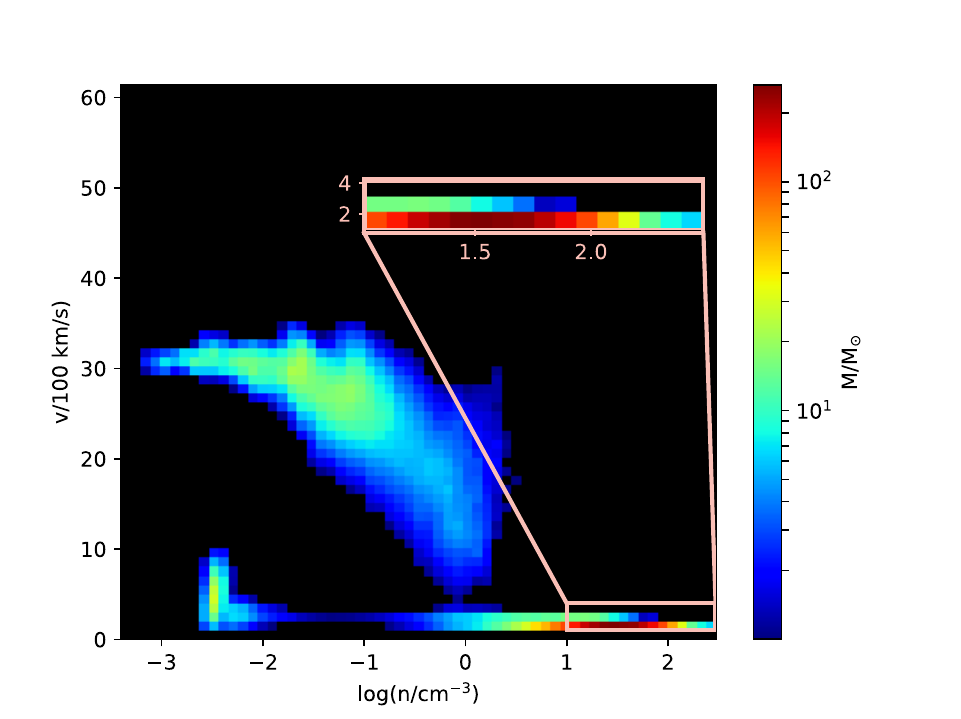}
\caption{The density-velocity map for \textit{f100n1500v} at 7 Myr. The little pink box shows the observed mean density and velocity range, and the larger pink box shows its zoom-in picture. The velocity is binned for every 100 km s$^{-1}$, so it can be conveniently read by counting the bins. The map shows the mass of every bin in solar mass, i.e., we can also estimate the mass of different components by counting the bins. The bins lower than 1 M$_\odot$ are suppressed.
\label{fig:vBnv}}
\end{figure}

\begin{figure}
\includegraphics[width=0.5\textwidth]{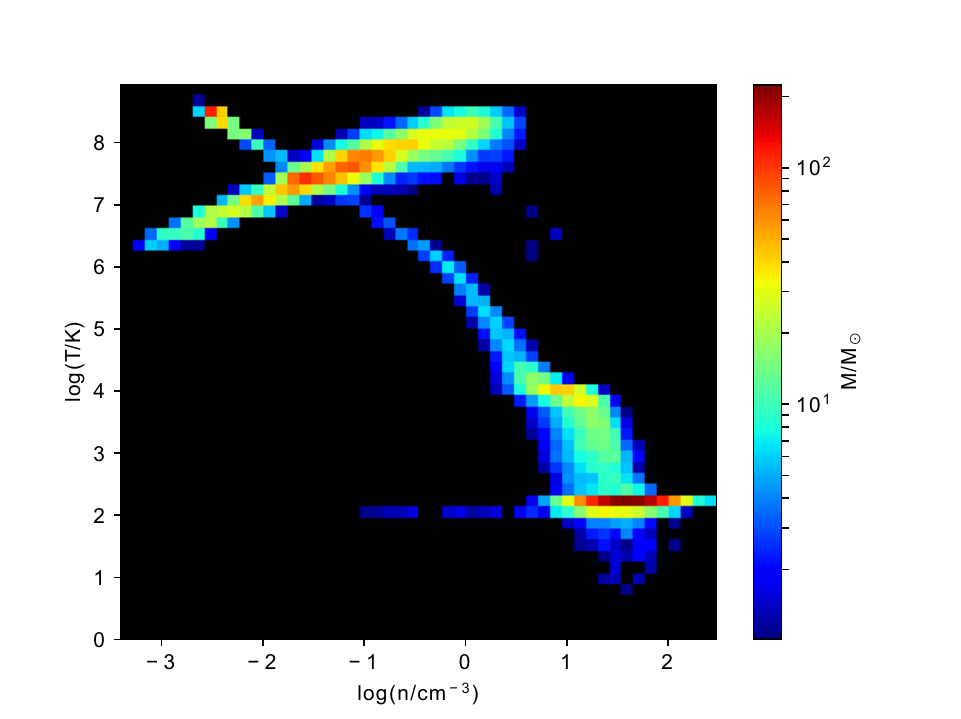}
\caption{The density-temperature maps of \textit{f100n1500v} at 7 Myr. The map shows the mass of every bin in solar mass, i.e., we can also estimate the mass of different components by counting the bins. The bins lower than 1 M$_\odot$ are suppressed.
\label{fig:vBnT}}
\end{figure}

The column density and velocity evolution of \textit{f100n1500v} is shown in Figure~\ref{fig:vB}, in which the clouds will reach 1 kpc at 7 Myr.
In the following text, we call the time, at which the results well match the observation, as the fiducial time.
At the early stage, the supernova shock wave would blow the initial MC to be a thin filament, because the central density of the cloud was much higher than the boundary.
The filamentary structures have also been investigated by \citet{BandaBarragan2016, 2023MNRAS.tmp.1171J}, who claim the filaments are only formed in magnetized environment and the cloud will crush to small clumps without magnetic field, consistent with our results.
When the peripheral low-density material was blown to higher latitude, the central dense core was being slowly accelerated.
Some pioneer high-velocity clumps broke away from the main cloud at 3 Myr, and run outside of the simulation box at 4 Myr.
At this stage, the main cloud became more irregular, but kept as one cluster.
After 7 Myr, the cloud would reach 1 kpc, a position consistent with the observation.
During the propagation of the cloud, the supernovae shock was always being reflected by the cloud and gradually produced stronger reverse shock.
This process leads to the obvious dividing line both for the density and velocity at 7 Myr.
The reverse shock could roughly balance the forward shock, as a result, the cloud acceleration rate largely decreased.

We also show the temperature and magnetic field evolution in Figure~\ref{fig:TvB}.
The outflow wind interact with the MCs, heating the surrounding ISM and compressing the magnetic field, while the central cores of the clouds still contain cool gas and low magnetic field at the early stage.
The shock wave from the supernovae can sweep the whole simulation box at $\sim$ 1 Myr and heat the ISM to high temperature.
However, with the receding of a part of the outflow wind at higher latitude, the magnetic field becomes much weaker.

Figure~\ref{fig:vB} \& \ref{fig:TvB} also show the starburst wind is not constant, especially at low latitude, because we adopt the random supernovae explosions in the simulations.
The ever-changing wind will significantly influence the evolution of the initial cloud.
However, the variation of the starburst wind is small at high latitude, where it can be taken as a constant wind.

To study whether the clouds at 7 Myr can be still taken as MCs with a velocity of $\sim$ 200 km $^{-1}$, we show the density-velocity and density-temperature maps in Figure~\ref{fig:vBnv} and \ref{fig:vBnT}.
At this moment, the simulation box contains three components: the clouds, ISM-dominated and SNR-dominated region, respectively corresponding to the lower right, lower left and central part of Figure~\ref{fig:vBnv}, and the lower right part, the central and the upper left band of Figure~\ref{fig:vBnT}.
Figure~\ref{fig:vBnT} is similar to the Figure 8 of \citet{2017ApJ...834..144S}, but we replace their constant wind with the simulated starburst wind.
As a result, Figure~\ref{fig:vBnT} contains the SNR-dominated region, i.e, the upper left band, which is absent in their work.
The clouds selected based on criteria of n$\geq$ 1 H cm$^{-3}$ and T$\leq$ 10$^4$ K have a total mass of $\sim$ 1500 M${\odot}$, while those selected based on criteria of n$\geq$ 10 H cm$^{-3}$, T$\leq$ 200 K, and $z \geq$800 pc are taken as molecular clouds and have a molecular mass of $\sim$ 850 M${\odot}$.
%The clouds have a total mass of $\sim$ 1100 M$_{\odot}$ (n$\geq$ 1 H cm$^{-3}$, T$\leq$ 10$^4$ K) and a molecular mass of $\sim$ 600 M$_{\odot}$ (n$\geq$ 10 H cm$^{-3}$, T$\leq$ 200 K, $z \geq$800 pc), both higher than the observation.
However, these clouds cover a region larger than the MW-C1 and MW-C1, and we should compare parameters at same scale.
If we choose the densest central clouds (diameter $\sim$30 pc, i.e., 60 cells) as the counterpart, the mass can better match the observation.
The clustering of the clouds is also considered in the estimation, in which some cells with appropriate density and temperature will still be excluded, if there is not any cloud cell within the surrounding 0.5 pc.
We also estimate the present mass-weighted mean velocity of $\sim$ 190 km s$^{-1}$ for all clouds (n$\geq$ 1 H cm$^{-3}$ and T$\leq$ 10$^4$ K), while the mean velocity over the past 7 Myr is $\sim$ 130 km s$^{-1}$, both a little lower than the observation.

In the vertical magnetic field, the clouds can propagate to 1 kpc without destruction, and still keep a considerable mass even larger than the observed HVMCs.
However, the mean velocity is a little smaller than the typical value.
To increase the velocity, a straightforward method is to increase the supernovae explosion frequency, but the frequency used in our work is already a little higher than the standard value.
In addition, it is unexpected that including the horizontal magnetic field can also increase the velocity, which will be illustrated in the next section.
Assuming a lower MCs or ISM density is also practical, so we test a case with a lower density of the initial MC in Section~\ref{subsec:rh}.
In summary, the fiducial run can indeed explain the acceleration of MCs at high latitude, while some features cannot be reproduced perfectly.

\subsection{The run with horizontal magnetic field}\label{subsec:hB}

\begin{figure*}
\includegraphics[width=\textwidth,height=\textheight,keepaspectratio]{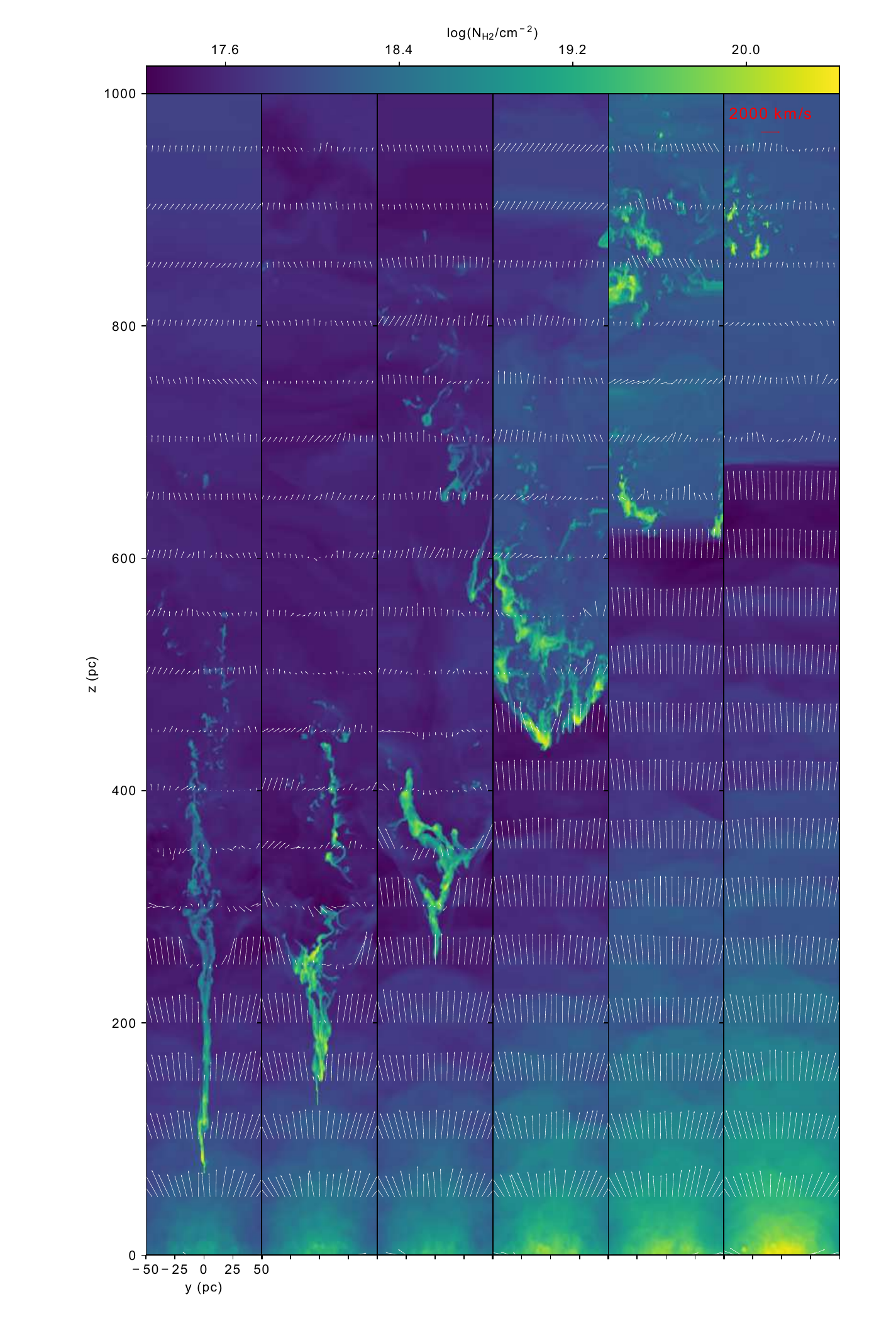}
\vskip-0.5cm
\caption{The column density maps of \textit{f100n1500h} between 1$\sim$6 Myr with a step of 1 Myr. The white arrows show the flow velocity in the slice through the $x$ = 0 pc, and the scale is shown at the upper right. The main cloud can also survive until it reaches 1 kpc , but it will lost much more mass than \textit{f100n1500v}.
\label{fig:hB}}
\end{figure*}

\begin{figure}
\includegraphics[width=0.5\textwidth]{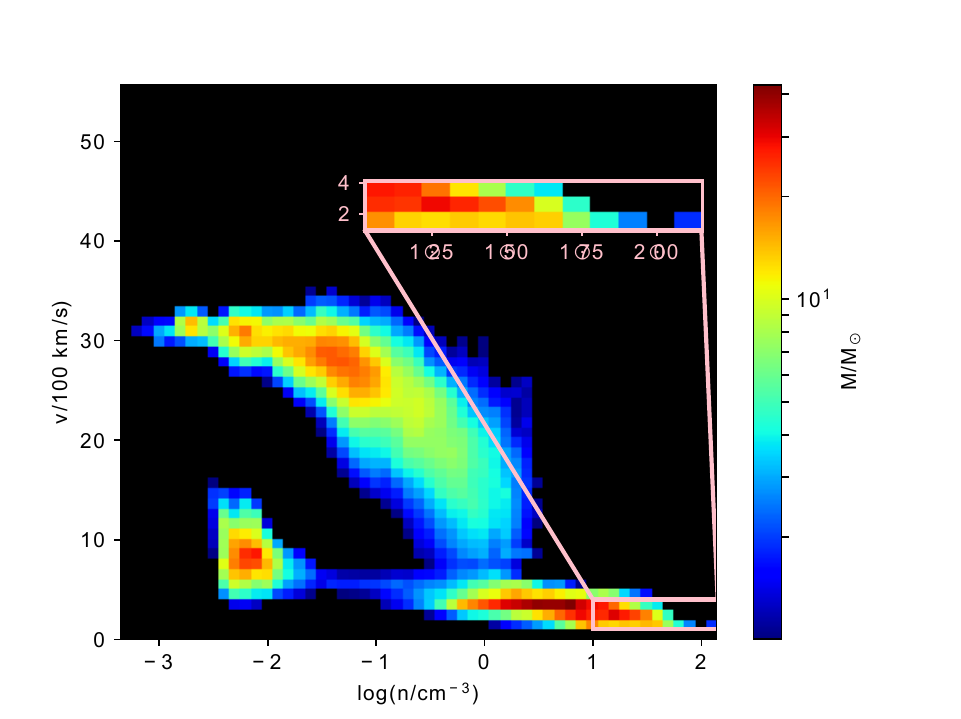}
\caption{The density-velocity map for \textit{f100n1500h} at 5 Myr. The little pink box shows the observed mean density and velocity range, and the larger pink box shows its zoom-in picture. The velocity is binned for every 100 km s$^{-1}$, so it can be conveniently read by counting the bins. The map shows the mass of every bin in solar mass, i.e., we can also estimate the mass of different components by counting the bins. The bins lower than 1 M$_\odot$ are suppressed.
\label{fig:hBnv}}
\end{figure}

We show the column density evolution of \textit{f100n1500h} in Figure~\ref{fig:hB}, while the density-velocity distribution at 5 Myr is shown in Figure~\ref{fig:hBnv}.
Similar to \textit{f100n1500v}, the MC was blown to be a thin filament initially, but gradually some gas was stripped.
At 2 Myr, a pioneer high-velocity clump separated from the main cloud, but run outside the simulation box at 3 Myr.
With the gas stripping, the MC showed a more irregular shape and finally crushed to several clumps.
These clumps have lower densities and higher velocities, but can be still taken as molecular clouds.
Especially, there is the second high-velocity clump separating from the main cloud after 4 Myr and reaching $\sim$ 1 kpc after 5 Myr, with a mass-weighted mean velocity of $\sim$ 340 km s$^{-1}$, a total mass of $\sim$ 700 M$_{\odot}$ (n$\geq$ 1 H cm$^{-3}$) and a molecular mass of $\sim$ 100 M$_{\odot}$ (n$\geq$ 10 H cm$^{-3}$).
The velocity is higher, but the masses are both lower than the observation's.

By comparing with \textit{f100n1500v}, we find a horizontal magnetic field can stimulate the acceleration and the crushing of the MCs, which is possibly caused by the magnetic tension force vertical to the Galactic plane, i.e., the magnetic draping, a ubiquitous mechanism already found in the launching of clouds \citep{2020ApJ...892...59C}.
The outflow wind can compress the MCs and the surrounding ISM, then amplify the local magnetic field, i.e., the magnetic tension.
The magnetic field can help to efficiently accelerate the MCs, while some MCs material will flow along the magnetic field, even run outside of the simulation box.
As a result, the MCs can be pushed to high velocity at high latitude, but lost much mass.
In addition, if the magnetic field includes more horizontal components, the clouds can be further dispersed at large scale, which can produce some smaller clouds than those in   \textit{f100n1500v}.
These clouds may be more similar to the observed MW-C1 and MW-C2.

In fact, the magnetic field in the Galactic center is complicated, while the vertical component is more important \citep{2013lsmf.book..215B, 2017JCAP...10..019C}.
At present, there is no a standard magnetic field model, so we test the two runs to study the influence of the magnetic direction on the simulation.
In terms of the two runs, a mixed magnetic field would likely better explain the observed properties of the HVMCs.

\subsection{The run without magnetic filed}\label{subsec:nB}

\begin{figure*}
\includegraphics[width=\textwidth,height=\textheight,keepaspectratio]{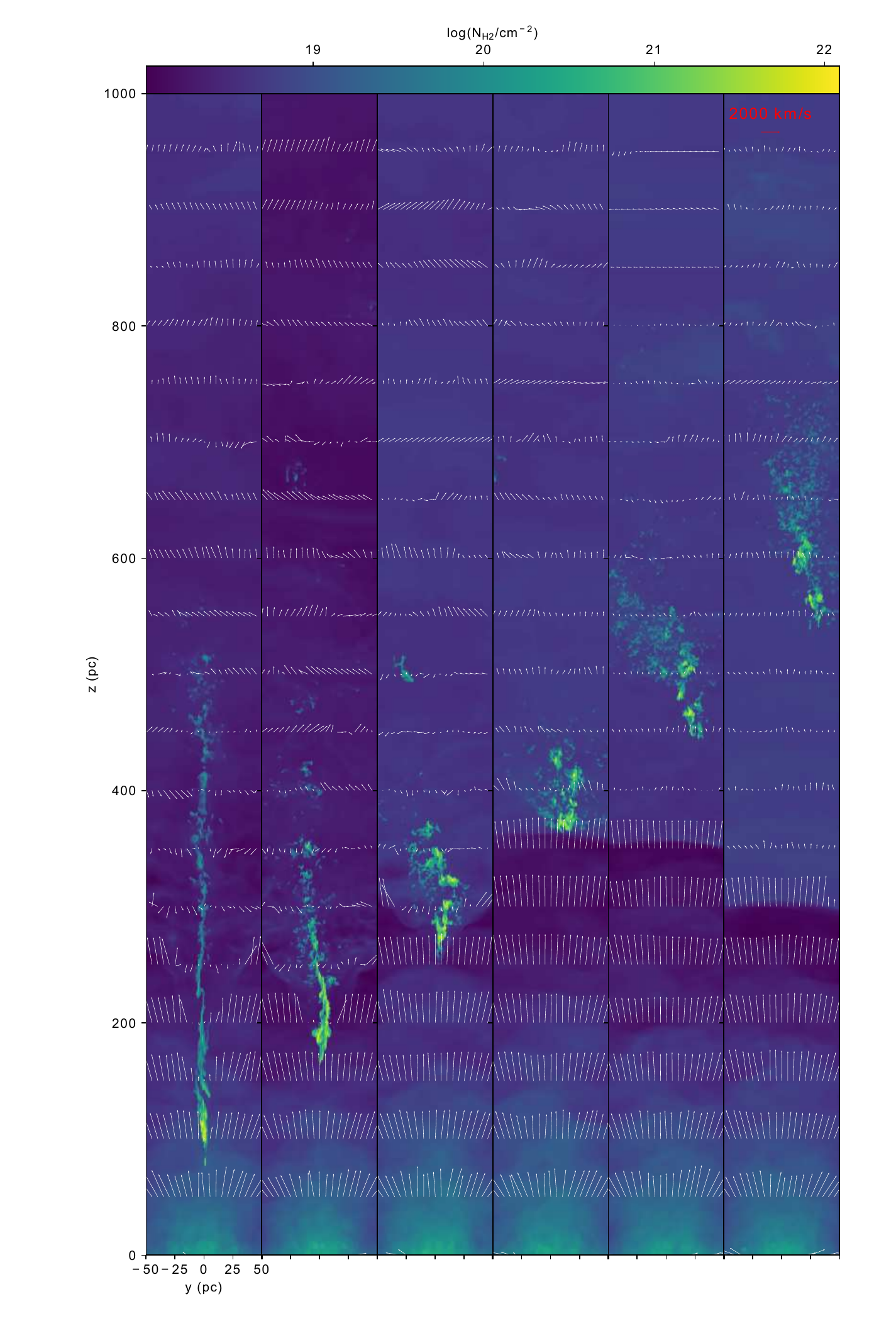}
\vskip-0.5cm
\caption{The $y$-$z$ column density maps of \textit{f100n1500n} between 1$\sim$6 Myr with a step of 1 Myr. The white arrows show the flow velocity in the slice through the $x$ = 0 pc, and the scale is shown at the upper right. The main cloud crushed quickly and cannot survive after 6 Myr.
\label{fig:nB}}
\end{figure*}

\begin{figure}
\includegraphics[width=0.5\textwidth]{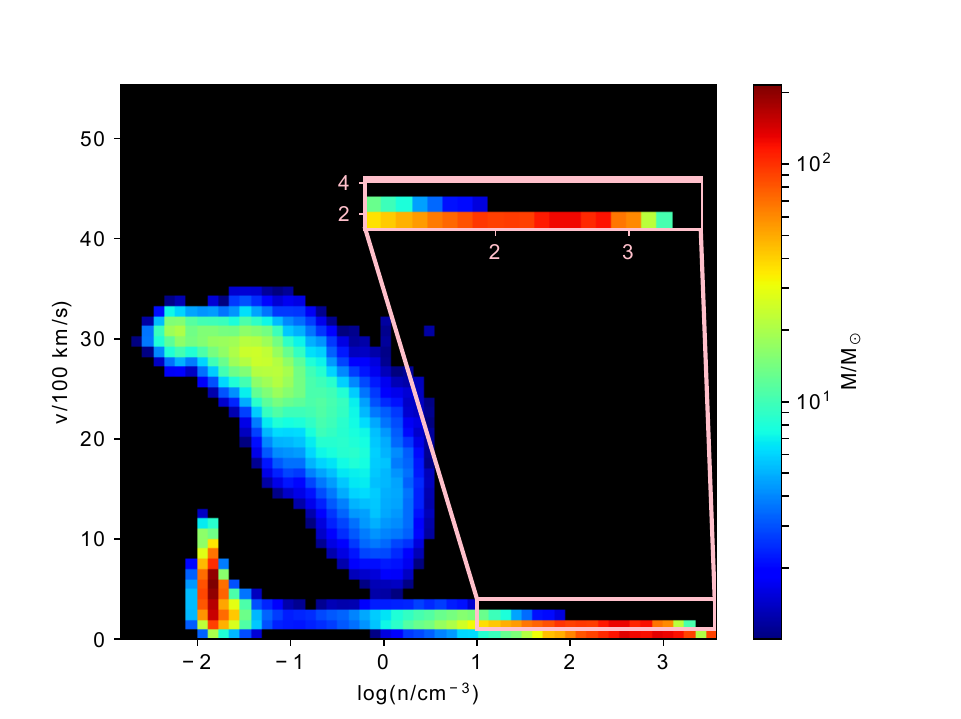}
\caption{The density-velocity map for \textit{f100n1500n} at 5 Myr. The little pink box shows the observed mean density and velocity range, and the larger pink box shows its zoom-in picture. The velocity is binned for every 100 km s$^{-1}$, so it can be conveniently read by counting the bins. The map shows the mass of every bin in solar mass, i.e., we can also estimate the mass of different components by counting the bins. The bins lower than 1 M$_\odot$ are suppressed.
\label{fig:nBnv}}
\end{figure}

We show the column density evolution of \textit{f100n1500n} in Figure~\ref{fig:nB}, and the density-velocity distribution after 5 Myr in Figure~\ref{fig:nBnv}.
There are no large clumps separation. Instead, lots of small clumps gradually diffuse from the main cloud, consistent with the simulation results of \citet{2017ApJ...834..144S}.
The main cloud is slower and will be depleted after 5 Myr, roughly consistent with the crushing time estimation, as a result, the MCs will not reach 1 kpc.
In other words, in comparison with \textit{f100n1500v} and \textit{f100n1500h}, the the magnetic field can indeed well protect the clouds.

However, Figure~\ref{fig:nB}  illustrates the densest regions are significantly  denser, and Figure~\ref{fig:nBnv} also shows there are more high-density clouds ($\rm~n~\geq~1000~H_2~cm^{-3}$)  than the runs with magnetic field, which indicates the magnetic field stimulates the destruction of the high-density clouds.
This effect may be attributed to the increased turbulence resulting from the presence of the magnetic field, facilitating a more efficient mixing of the MCs and interstellar medium (ISM).
As a consequence, high-density clouds share material with low-density regions.
Additionally, the clumps grow larger and exhibit prolonged survival but possess lower densities.
Consequently, the local density of the clouds is diminished in the presence of a magnetic field.

In conclusion, the magnetic field plays a crucial role in the formation of HVMCs.
However, it should be noted that the magnetic field does not always increase the density of MCs and can also disperse some of the densest MCs at smaller scales.

\subsection{The run with lower density and lower explosion frequency}\label{subsec:rh}

\begin{figure*}
\includegraphics[width=\textwidth,height=\textheight,keepaspectratio]{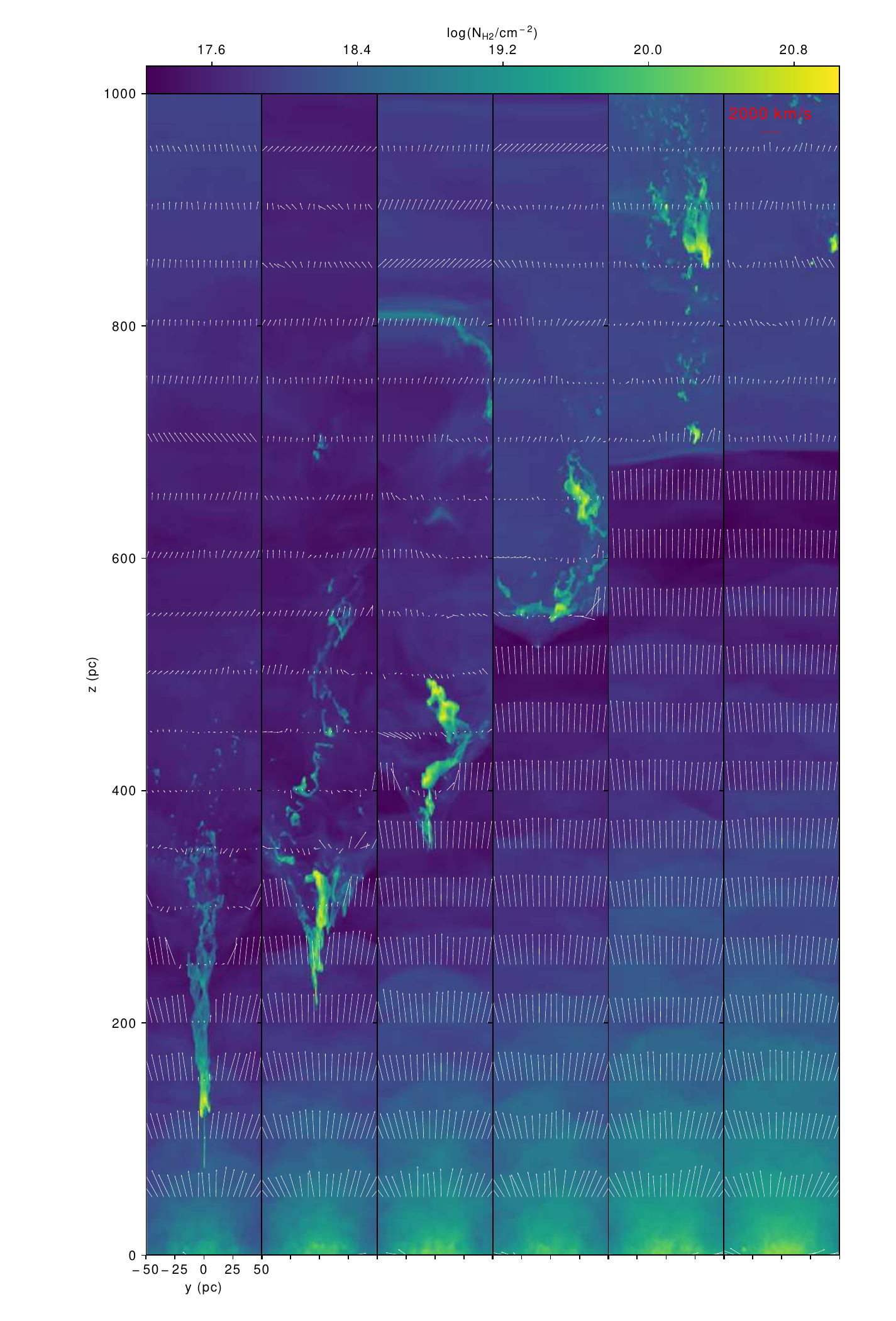}
\vskip-0.5cm
\caption{The $y$-$z$ column density maps of \textit{f100n1000v} between 1$\sim$6 Myr with a step of 1 Myr. The white arrows show the flow velocity in the slice through the $x$ = 0 pc, and the scale is shown at the upper right. The main cloud can indeed survive until it reaches 1 kpc at 5 Myr, though it will lost a large amount mass.
\label{fig:rh}}
\end{figure*}

\begin{figure}
\includegraphics[width=0.5\textwidth]{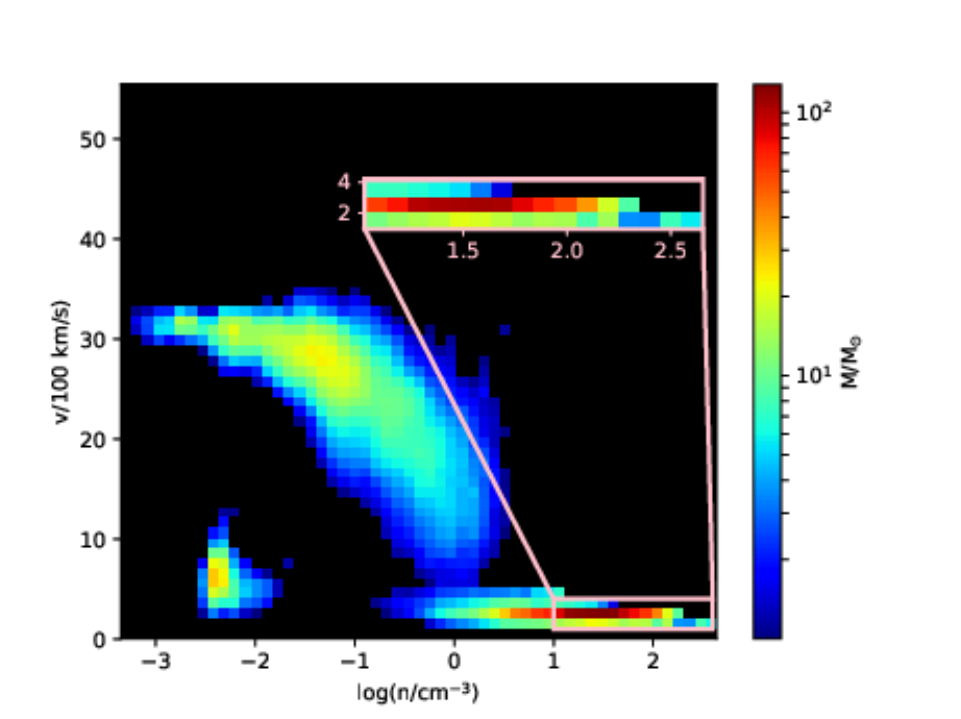}
\caption{The density-velocity map for \textit{f100n1000v} at 5 Myr. The little pink box shows the observed mean density and velocity range, and the larger pink box shows its zoom-in picture. The velocity is binned for every 100 km s$^{-1}$, so it can be conveniently read by counting the bins. The map shows the mass of every bin in solar mass, i.e., we can also estimate the mass of different components by counting the bins. The bins lower than 1 M$_\odot$ are suppressed.
\label{fig:rhnv}}
\end{figure}

We show the results of \textit{f100n1000v} in Figure~\ref{fig:rh} and Figure~\ref{fig:rhnv}.
The initial cloud was also blown to a filament, but a little wider than previous runs.
A large amount of gas were stripped at 2 Myr, and dissipated at 3 Myr.
Then the main cloud was divided to two clouds, in which the faster one almost reached 1 kpc at 5 Myr, but the left one gradually disappeared.
After 5 Myr, the simulation box has a total mass of $\sim$ 1100 M$_{\odot}$ (n$\geq$ 1 H cm$^{-3}$, T$\leq$ 10$^4$ K) and a molecular mass of $\sim$ 500 M$_{\odot}$ (n$\geq$ 10 H cm$^{-3}$, T$\leq$ 200 K, $z \geq$800 pc), roughly consistent with the MW-C2.
The mass-weighted mean velocity is $\sim$ 290 km s$^{-1}$, also similar to the observation.
However, the diameter of the whole cloud is larger than the observation, so the density is lower.

By comparing with \textit{f100n1500v}, we can estimate a central density between 1000 and 1500 cm$^{-3}$ for the initial cloud.
Meanwhile, if the magnetic field includes more horizontal components, the clouds can be further dispersed to some small clouds.
In other words, if \textit{f100n1500v} uses a lower central density and more horizontal magnetic field, it can better match the observation.
However, the primary focus of this study is to investigate whether the MCs can be accelerated to high velocities at high latitudes, and the current findings adequately address this inquiry.
Moreover, it is important to note that the parameters for MW-C1 and MW-C2 are only approximations derived from a simplified biconical wind model, and the completeness of the HVMCs sample remains uncertain.
There are only two detected HVMCs in the Galactic center, so the main feature of HVMCs is actually still ambiguous.
As a result, conducting an exhaustive search of the parameter space is unnecessary at this stage.

\begin{figure*}
\includegraphics[width=\textwidth,height=\textheight,keepaspectratio]{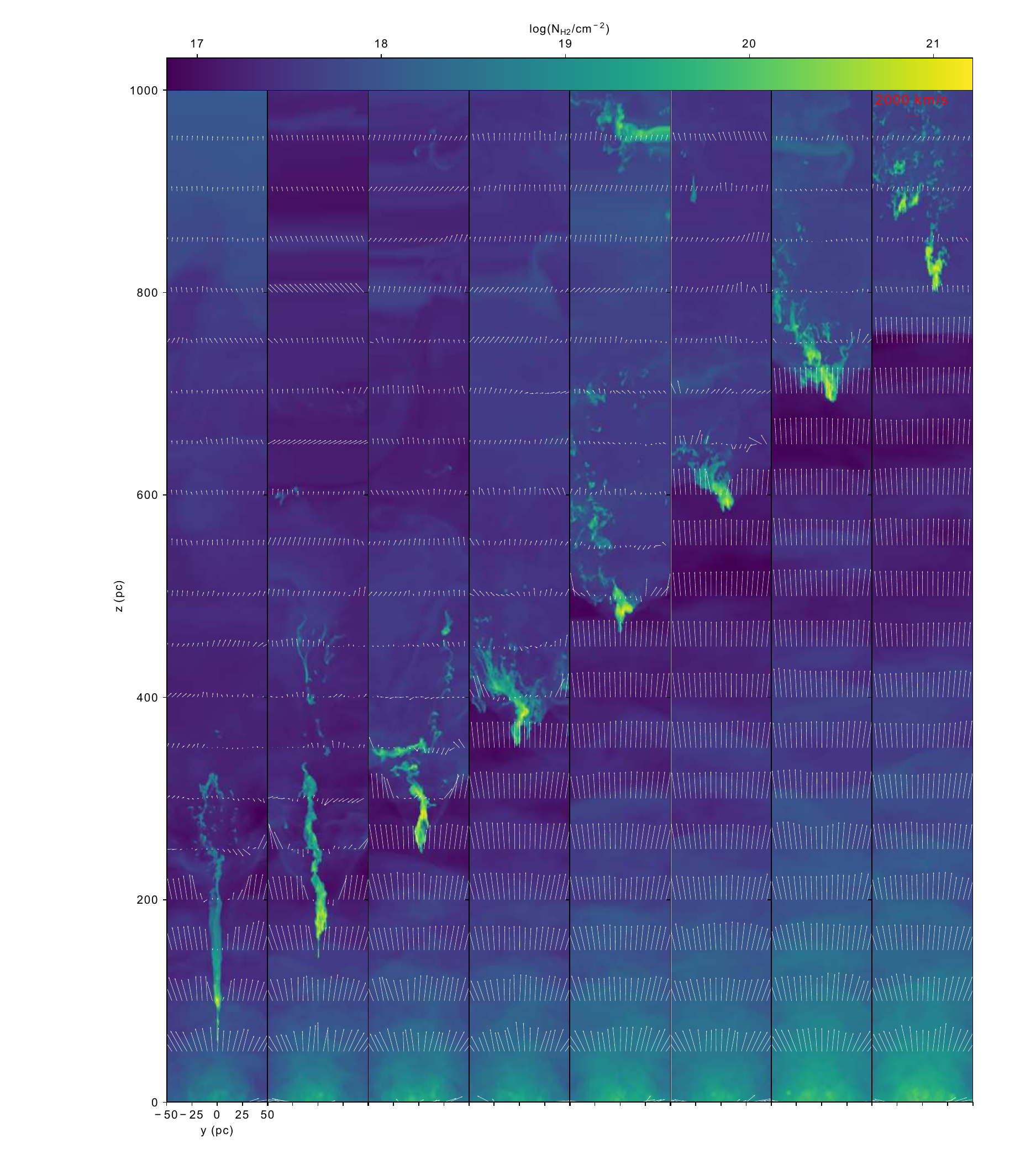}
\vskip-0.5cm
\caption{The $y$-$z$ column density maps of \textit{f200n1000v} between 1$\sim$8 Myr with a step of 1 Myr. The white arrows show the flow velocity in the slice through the $x$ = 0 pc, and the scale is shown at the upper right. The main cloud can indeed survive until it reaches 1 kpc at 8 Myr, though it will lost a large amount mass.
\label{fig:f2}}
\end{figure*}

\begin{figure}
\includegraphics[width=0.5\textwidth]{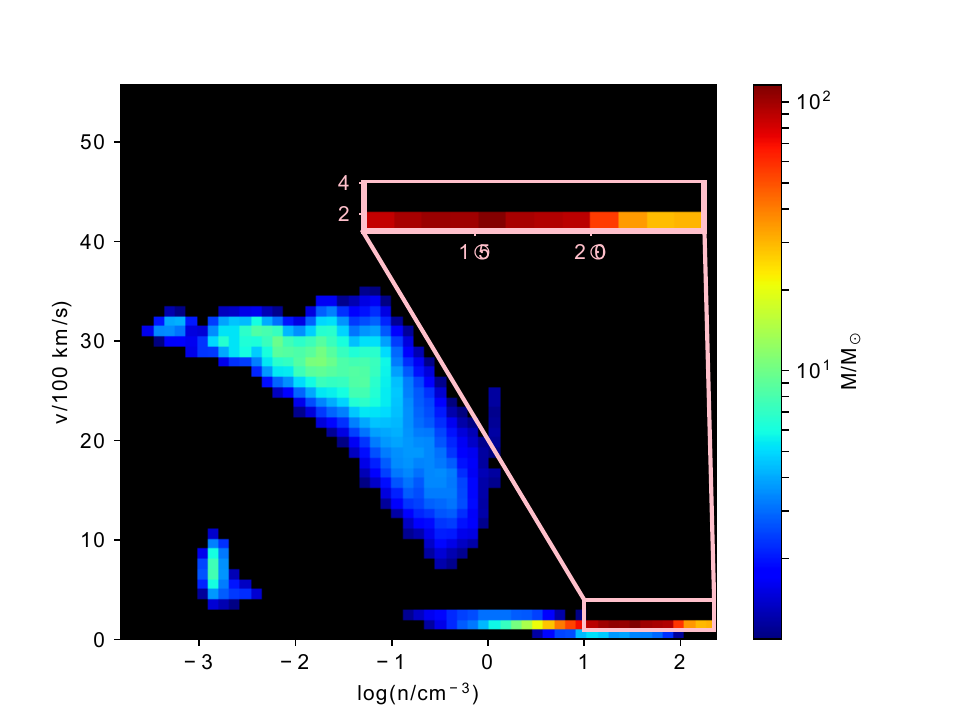}
\caption{The density-velocity map for \textit{f200n1000v} at 8 Myr. The little pink box shows the observed mean density and velocity range, and the larger pink box shows its zoom-in picture. The velocity is binned for every 100 km s$^{-1}$, so it can be conveniently read by counting the bins. The map shows the mass of every bin in solar mass, i.e., we can also estimate the mass of different components by counting the bins. The bins lower than 1 M$_\odot$ are suppressed.
\label{fig:f2nv}}
\end{figure}

The results of \textit{f200n1000v} are shown in Figure~\ref{fig:f2} and Figure~\ref{fig:f2nv}.
The gas were gradually stripped, but the cloud can still survive to reach 1 kpc at 8 Myr.
At 8 Myr, the simulation box has a total clouds mass of $\sim$ 640 M$_{\odot}$ ( n$\geq$ 1 H cm$^{-3}$, T$\leq$ 10$^4$ K), a molecular mass of $\sim$ 400 M$_{\odot}$ (n$\geq$ 10 H cm$^{-3}$, T$\leq$ 200 K, $z \geq$800 pc) and a mass-weighted mean velocity is $\sim$ 180 km s$^{-1}$, roughly consistent with MW-C1, which indicates the cloud can also well survive, even if the explosion frequency of the supernovae is lower than $10\rm~kyr^{-1}$.
In fact, the explosion frequency should vary with the evolution of the cloud, and the features of resultant clouds are also dependent on the variation.

\begin{figure}
\includegraphics[width=0.5\textwidth]{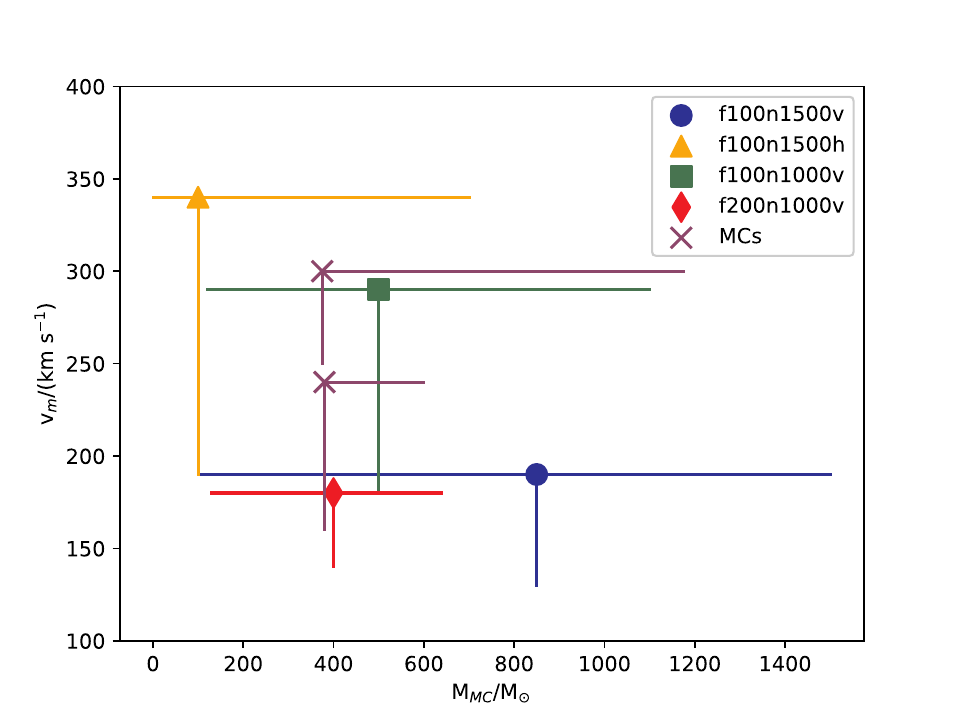}
\caption{Visualization of the Mass-Velocity Relation presented in Table~\ref{table:result}. For the simulation results, the central points depict the molecular mass and mass-weighted velocity. The total mass and the MCs mass on strict criteria (see text for details) serves as the upper limit and lower limit, while the velocity on another strict criteria represents the lower limit of the velocity. For MW-C1 and MW-C2, the central points display the molecular mass and outflow velocity, while the total mass serves as the upper limit and the velocity along the line of sight represents the lower limit.
\label{fig:para}}
\end{figure}

We summarize all results in Table \ref{table:result} and visualize it in Figure \ref{fig:para}, which will be further discussed in Section \ref{sec:dis}.
The criteria (n$\geq$ 10 H cm$^{-3}$, T$\leq$ 200 K, $z \geq$800 pc) used to choose the molecular components is not always reasonable, and some hydrogen atoms can also survive on such a criteria.
We here show the results with a strict criteria (n$\geq$ 100 H cm$^{-3}$, T$\leq$ 150 K, $z \geq$800 pc) for an error estimation.
It is actually difficult to accurately estimate the realistic velocity of the clouds based on the current observation.
We take the velocity along the sightlines as the lower limit, and the outflow velocity as the standard velocity.
The outflow velocity is estimated based on a biconical model, which is also an important reference for our simulations, so we use the outflow velocity to directly compare with the simulations.

\begin{table*}
\center
\caption{The resultant parameters at the fiducial times}
\label{table:result}
\begin{threeparttable}
\begin{tabular}{ccccccccc} % four columns, alignment for each
\hline
Run \& MCs & $M_{\rm tot}$ & $M_{\rm MC1}$ & $v_{\rm m1}$ & $M_{\rm MC2}$ & $v_{\rm m2}$  & $t_{\rm f}$ & $v_{\rm LoS}$ & $v_{\rm outflow}$ \\
\hline
(1) & (2) & (3) & (4) & (5) & (6) & (7) & (8) & (9)\\
\hline
\textit{f100n1500v}   & 1500 & 850 & 190 & 100 & 130 & 7\\
\textit{f100n1500h}   & 700  & 100 & 340 & 2   & 190 & 5\\
\textit{f100n1000v}   & 1100 & 500 & 290 & 120 & 170 & 5\\
\textit{f200n1000v}   & 640  & 400 & 180 & 130 & 140 & 8\\
\textit{MW-C1}        & 600  & 380 &     &     &     &     &  160 & 240 \\
\textit{MW-C2}        & 1175 & 375 &     &     &     &     &  250 & 300 \\
\hline
\end{tabular}
\begin{tablenotes}
\footnotesize
\item (1) The runs and MCs. (2) The total mass both including atoms and molecules, in units of M$_{\odot}$, and the selection criterion is n$\geq$ 1 H cm$^{-3}$, T$\leq$ 10$^4$ K. (3) The molecular mass above 800 pc, in units of M$_{\odot}$, and the selection criterion is n$\geq$ 10 H cm$^{-3}$, T$\leq$ 200 K, $z \geq$800 pc. (4) The simulated mass-weighted mean velocity, in units of km s$^{-1}$. (5) The molecular mass above 800 pc, in units of M$_{\odot}$, and the strict selection criterion is n$\geq$ 100 H cm$^{-3}$, T$\leq$ 150 K, $z \geq$800 pc. (6) The simulated mass-weighted mean velocity on the strict criteria, in units of km s$^{-1}$. (7) The fiducial times, in units of Myr. (8) The velocity along the line of sight, in units of km s$^{-1}$. (9) The outflow velocity estimated based on a biconical model, in units of km s$^{-1}$.
\end{tablenotes}
\end{threeparttable}
\end{table*}

\section{Discussion}\label{sec:dis}
In the preceding sections, we have presented 3D simulations that illustrate the long-term hydrodynamic evolution of MCs propelled by subsequent supernova explosions. These simulations incorporate simplified, yet sufficiently realistic physical conditions of both the MCs and the surrounding environment. The first three simulation runs, which represent the evolution with vertical, horizontal, and no magnetic field, exhibit varying degrees of success and shortcomings in replicating the primary observed characteristics of MW-C1 and MW-C2. The last two runs show the simulations work well in a wide parameter space. In this section, we analyze the outcomes of these simulations and discuss their implications for our comprehension of the enigmatic ecosystem in the Galactic center.

\subsection{Formation and evolution of the HVMCs}
\label{subsec:form}
Table \ref{table:result} demonstrates that the total mass of the four runs with magnetic field aligns with the observed HVMCs, while \textit{f100n1500h} exhibits a lower molecular mass and higher velocity.
Figure \ref{fig:para} clearly indicates that \textit{f100n1000v} provides the closest match to the two HVMCs, though the other two runs also show rough consistency with the observations.
However, the key point we want to make is that the HVMCs can indeed be accelerated to high velocities without disruption, which is also reflected by the other three runs.
We can assume the position of the central point shown in Figure \ref{fig:para} can be interpolated accordingly, if we change one of the parameters, such as the direction of the magnetic field, the supernovae explosion frequency and the density of the initial cloud, based on which we can roughly estimate the dependence of their positions on these parameters.
For example, by comparing \textit{f100n1500v} with \textit{f100n1500h} and drawing a line between the two central points, we can expect a point will be located between the two MCs, when we only modify the direction of the magnetic field.
Similarly, we can get a cloud with higher density and velocity than the two MCs by properly increase the supernovae frequency or decrease the initial cloud density.

All four runs with magnetic field can reproduce the HVMCs, so the HVMCs can be indeed formed by the acceleration of the starburst in the Galactic center.
On the other hand, these results indicate the magnetic field is important and the MCs can well survive the shock of supernovae even at a scale of $\sim$ 1 kpc.
\citet{2017MNRAS.468.4801Z} claim the cold gas with temperature of 10$^2$ $\sim$ 10$^4$ K cannot survive a hot Galactic wind, but they neglect the magnetic field and pay more attention to study the process at larger scale, which will not conflict with our results.
Of course, in our simulations, there are also some features inconsistent with the observations, so we will try to clarify them in this section.

\begin{figure}
\includegraphics[width=0.5\textwidth]{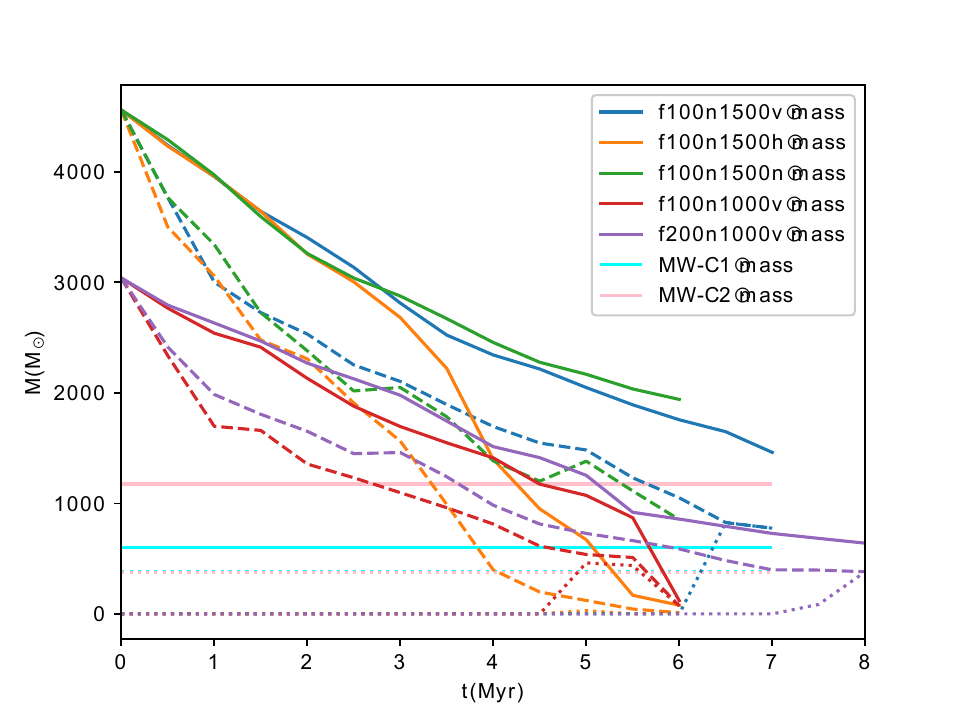}
\caption{The mass evolution of HVMCs. The solid, dashed and dotted lines respectively show the corresponding the total mass, the total molecular mass and the molecular mass with a latitude higher than 800 pc.
\label{fig:m}}
\end{figure}

To better study the evolution of the HVMCs, we show the mass evolution of all runs in Figure~\ref{fig:m}.
The criterion for distinguishing between various components follows the description presented in Section~\ref{subsec:vB}.
There are three kinds of mass, the total gas mass (atoms + molecules), the total molecular mass and the mass of molecular gas with a latitude higher than 800 pc.
For simplicity, we take the last one as the molecular mass of the HVMCs.

This analysis of the mass evolution of the HVMCs shows that the total gas and molecular mass of all five runs gradually decrease over time due to ionization, stripping by the hot wind and outflows from the simulation box.
However, \textit{f100n1500h} and \textit{f100n1000v} show a rapid declination respectively after 3.5 Myr and 5.5 Myr.
For \textit{f100n1500h}, this is caused by the dissipation of the pioneer high-velocity clump  which quickly diffuse and run out of the left and right edges of the simulation box along the horizontal magnetic field.
Similar to \textit{f100n1500h}, \textit{f100n1000v} also has a high-velocity clump running out of the simulation box, but from the upper edge after 5.5 Myr.
As for the total molecular clouds, they are stripped and dissociated rapidly at the beginning, and maintain a steady decrement.
At last, \textit{f100n1500h} and \textit{f100n1000v} lost most of molecular gas after 6 Myr, while a large amount still survive in \textit{f100n1500v}, \textit{f100n1500n} and \textit{f100n1000v}.
In \textit{f100n1500n}, the clouds, almost totally crushed after 5 Myr, cannot approach 800 pc, so they are impossible to form the HVMCs.

In \textit{f100n1500v}, the total gas mass and HVMCs mass are much higher than MC-C2 after 7 Myr, while in \textit{f200n1000v},  they are comparable to MC-C1 at 8 Myr.
At this stage, the total molecular mass istotally composed of the HVMCs mass, so all of the molecular components have propagate beyond 800 pc.
In \textit{f100n1500v}, the clouds have lower-velocity and larger volume than the observed HVMCs, but the mean density is similar.
Therefore, if the clouds crushed as some higher-velocity small clouds similar to the observed HVMCs, this run can better match the observation.
It happens that the clouds will diffuse to be some small clumps with a mass-weighted mean velocity of $\sim$ 340 km s$^{-1}$ in \textit{f100n1500h}, though the velocity becomes a little higher than the observations.
Therefore, it is natural to expect a magnetic field including both vertical and horizontal components, will help to produce the better-matched HVMCs in the simulation. Such a configuration is actually more reasonable for the real magnetic field in the Galactic center,
A general model for the whole Milky Way also shows the magnetic field is parallel to the Galactic plane at lower latitude and gradually tend to be perpendicular at higher latitude \citep{2017JCAP...10..019C}, so the expectation is sensible.
In addition, if a higher resolution (4 times) is applied in the simulation, the clouds will also crush to be smaller clumps \citep{2017ApJ...834..144S, 2020MNRAS.492.1970G}, of which velocity and total mass are similar to those in the lower resolution.
Therefore, it will be more consistent with the observations, since MW-C1 and MW-C2 are both smaller than the clouds produced in \textit{f100n1500v}, \textit{f100n1000v} and \textit{f100n1500h}.
In other words, the resolution used in our work is adequate to explain the formation of HVMCs, if we do not take the volume of the HVMCs as an essential feature.
Of course, using a low resolution, the simulations cannot accurately describe the instability and the mixing between the cold gas and the hot wind, which may stimulate the crushing of clouds, but the advection of hot high-enthalpy gas into the mixing layer actually can result in growth and acceleration of the cold phase \citep{2020ApJ...894L..24F}.

The observations show many HVCs distributed over a large latitude from $\sim$ 100 pc to $\sim$ 10 kpc \citep{2018ApJ...855...33D, 2020ApJ...888...51L, 2022MNRAS.513.3228L}, though most of the HVCs are located in the lower 2 kpc.
In our simulation, we only consider the starburst happening in a small region and include only one initial cloud, which limits the number of HVCs formed in the simulation box. However, the main focus of our work is to investigate the formation mechanism of HVMCs, rather than reproducing the exact number and distribution of observed HVCs. The fact that we can reproduce the key features of HVMCs observed in the Milky Way, such as their high velocity and high density, suggests that our proposed formation mechanism is plausible and can contribute to the understanding of the origin of HVCs in general. Further studies including more initial clouds and considering the starburst happening over a larger region would be needed to fully reproduce the observed distribution of HVCs.

The MCs in the run without magnetic field will be crushed in a short term, so the magnetic field is essential for the formation of the HVMCs.
The magnetic field can wrap and protect the MCs, a mechanism named as the magnetic draping, which is significant at a large scale range, from the small scale of comets to the large scale of galaxy clusters \citep{1986Natur.321..288R, 1996ApJ...465..800J, 2006Icar..182..464B, 2008ApJ...677..993D}. Therefore, it is possibly contributed to the survival of our HVMCs.
Nevertheless, Figure \ref{fig:TvB} shows the magnetic field surrounding the cold clouds is chaotic and does not well wrap the clouds, and our zoom-in check also shows same results, which is possibly caused by the low resolution and the wrapping is only obvious at much smaller scales.
\citet{2023MNRAS.tmp.1171J} try to study the survival of HVCs in the Galactic halo, and claim that magnetic fields suppress hydrodynamic instabilities and the growth of small-scale structures, which is also responsible for the protection of the HVMCs in our simulations.
In addition, the direction of the magnetic field can also influence the evolution of the HVMCs, which can be read from Figure \ref{fig:vB}, \ref{fig:hB} and \ref{fig:nB}.
In a vertical magnetic field, the clouds can keep high density and propagate to high latitude.
\citet{2020MNRAS.499.4261S} also conclude that the vertical magnetic field can well protect a cold cloud, but the cloud they used actually has a temperature of 10$^4$ K and a density of 0.1 cm$^{-3}$, totally different from the parameters used in our simulations.
In a horizontal magnetic field, the clouds will lose an amount of mass, but still can propagate to high latitude without crushing.
If there is not magnetic field, the clouds cannot propagate to high latitude.
The importance of direction is also discussed by \citet{2014MNRAS.444..971A} \& \citet{2020ApJ...892...59C}, though the properties of clouds, winds, magnetic field and ISM they used are different from ours.

Additionally, our simulations consistently demonstrate that the reverse shock generated by the interaction between the clouds and the Galactic wind effectively balances the forward shock at later stages. As the clouds propagate, the forward shock of the Galactic wind encounters resistance from the clouds, leading to the gradual formation of stronger reverse shocks. This phenomenon is clearly observed in the density-velocity distribution plots presented in Figure \ref{fig:vB, fig:hB, fig:nB, fig:rh} and \ref{fig:f2}. It is expected that at this late stage, the clouds have attained their maximum velocity within the framework of our model, and further acceleration becomes inefficient. Furthermore, we note that the star formation rate (SFR) employed in our model represents an upper limit within reasonable estimations, ensuring that the supernova explosion frequency is also maximized. Among the runs, \textit{f100n1500h} stands out with the highest velocity exceeding 400 km s$^{-1}$, although it should be noted that the assumption of a complete horizontal magnetic field in the Galactic center is not physically realistic. Thus, if our model accurately captures the physics, we predict that the maximum velocity attainable by the HVMCs would be approximately 400 km s$^{-1}$.

Overall, the simulation results provide a promising framework for explaining the formation of HVMCs and their connection to HVCs. The HVMCs can indeed originate from a starburst in the Galactic center, which is reasonable in a large parameter space.
The magnetic field can protect the MCs and contribute to the acceleration of MCs, but the acceleration of MCs is limited at high latitdue.
However, there are still many uncertainties and complexities involved in the process, such as the role of magnetic fields, the effects of different initial conditions, and the possible interactions with other structures in the Galactic center. Therefore, further investigations are needed to refine and extend the current model, and to test its validity against more detailed observations and simulations.

\subsection{The metallicity of HVCs}
\label{subsec:metal}
The formation of HVMCs is tightly associated with the HVCs', but the origin of HVCs is also ambiguous.
The HVCs are usually defined as the interstellar gas clouds that moving at speeds substantially different (up to several hundreds km s$^{-1}$) to the rotation of the disk of the Milky Way, and they are mostly  distributed in the whole Galactic halo.
Most of them have lower metallicity than what we find in the disk, so they may come from the Galactic halo or intergalactic medium.
However, some of them, especially in the Fermi bubbles, have much higher metallicity, so they may be ejected from the Galactic disk.
The HVCs in the Fermi bubbles are usually called as FB HVCs, which will be primarily discussed in this section.

It has been suggested that the HVCs are composed of diffuse inflowing gas and collimated outflowing material, which are likely manifestations of a galaxy-wide gas cycle triggered by stellar feedback, known as the galactic fountain \citep{2020ApJ...898..148L, 2022MNRAS.515.4176M}.
The feedback and the interaction with surrounding galaxies both influence the material cycle in our Milky Way, in which, most of the FB HVCs should be taken as a part of the collimated outflow \citep{2015ApJ...799L...7F, Bordoloi2017, 2020ApJ...898..128A}, because the stellar activity in the Galactic center is stronger than the disk.
However, \citet{2022NatAs...6..968A} found the FB HVCs have a wide range of metallicities from $\le$ 0.2 of solar to $\sim$ 3.2 $Z_{\odot}$, thus the gas from the halo may also mix with the local ISM and ejecta from the disk.
The supersolar metallicity of $\sim$ 3.2 $Z_{\odot}$ implies that the HVCs are initially metal-rich, or there is a metal-enrichment process during the acceleration of the HVCs, since the Galactic ISM metallicity is usually $\sim$ 1 solar \citep{2021ApJS..252...22Z}.
Therefore, it is convenient to assume the FB HVCs with high metallicity are formed by the driven of many sequential supernovae explosions which can simultaneously accelerate the clouds and provide heavy elements, a process also possibly happening in other galaxies \citep{2019MNRAS.482.1304E}.
The SMBH activity may also drive the HVCs, but a metal-enrichment process, i.e., the supernovae explosions, is always necessary.

\begin{figure}
\includegraphics[width=0.5\textwidth]{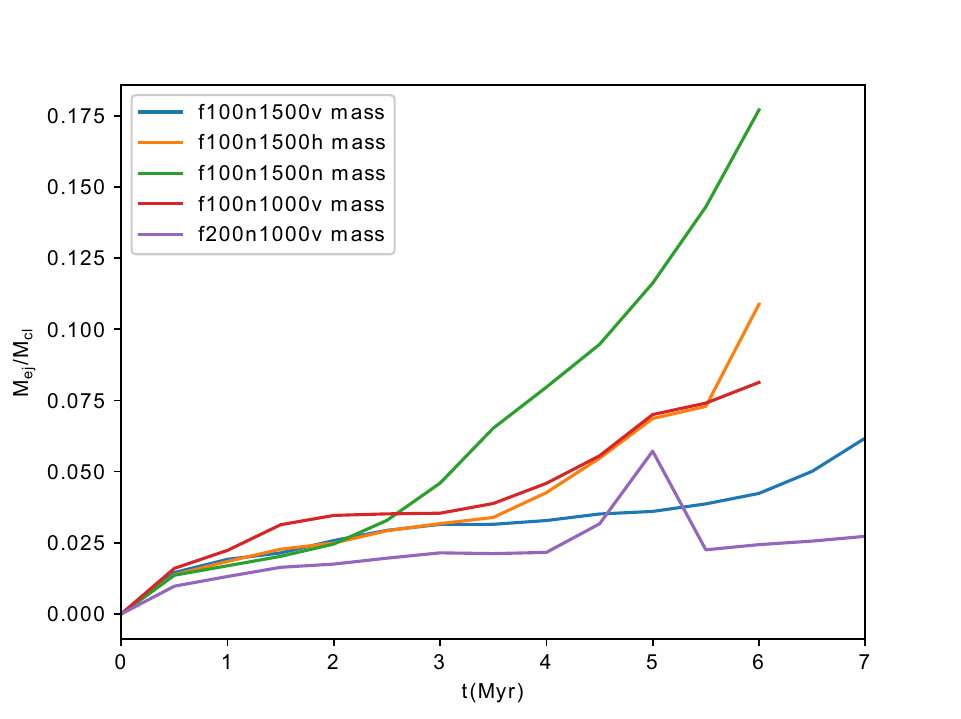}
\caption{The ratio of the ejecta mass to the total clouds mass in the simulation box.
\label{fig:metal}}
\end{figure}

The origin of HVMCs is likely analogous to FB HVCs, but this has yet to be confirmed due to the lack of information about their metallicity. To investigate this further, we examined the ratio of ejecta mass to cloud mass in our simulation, as shown in Figure~\ref{fig:metal}. The ratio generally increases over time for all runs, but there is a peak at 5 Myr for \textit{f2001000v}, which may be due to the low-metallicity cloud material flowing out of the simulation box.
In \textit{f100n1500n}, the clouds contain more ejecta material since they are slow, resulting in a more efficient mixture. Assuming an initial cloud metallicity of 1 $Z_{\odot}$ and a supernova ejecta metallicity of 6 $Z_{\odot}$, a standard ratio of 0.1 would yield a final cloud metallicity of 1.5 $Z_{\odot}$, still lower than the observed 3.2 $Z_{\odot}$ in some FB HVCs \citep{2022NatAs...6..968A}.
This suggests that the initial clouds were possibly already metal-rich before being driven to become HVCs. While SMBH activity may also drive HVCs, a metal-enrichment process such as supernova explosions is possibly necessary to explain the high metallicity of some FB HVCs.

If the model is correct, the role of HVCs in the galaxy-wide gas cycle can be understood. The low-metallicity HVCs originating from the halo or intergalactic medium are pulled by the gravitational potential of the Milky Way and surrounding galaxies, while high-metallicity HVCs are driven by galactic fountains that are energized by supernovae explosions in our Milky Way or the SMBH in the Galactic center. The FB HVCs consist of both types of HVCs, but the HVMCs embedded in FB HVCs should be driven by the fountains, which could be further confirmed by future metallicity analysis based on new ultraviolet absorption observations.

\subsection{The relation between HVMCs and feedback relics}
\label{subsec:rel}
It is interesting to ask whether the HVMCs have a causal relation with the radio bubbles \citep{2019Natur.573..235H} and X-ray chimneys \citep{2019Natur.567..347P} found on smaller scales, or the Fermi bubbles \citep{2010ApJ...724.1044S} and eROSITA bubbles \citep{Predehl2020} found on much larger scales.
We note that the age of the HVMCs inferred from our simulations is a few Myr, roughly consistent with the dynamical timescale of a few Myr for both of the radio bubbles and the Fermi bubbles originally suggested by \citet{2019Natur.573..235H} and \citet{2013MNRAS.436.2734Y}, respectively.
However, their timescales actually have not been resolved, the radio bubbles may be younger \citep[330 kyr]{2021ApJ...913...68Z} and the Fermi bubbles may be much older \citep[1 Gyr]{2011PhRvL.106j1102C}.
In particular, \citet{2019Natur.573..235H}'s estimation was based on the assumption of a constant expansion velocity of the bubbles, which is implausible, hence a shorter timescale is expected.
In the context of the supernova-based model for the origin of the radio bubbles/chimneys \citep{2021ApJ...913...68Z}, the radio bubbles would be a dynamically younger and independent structure simply evolving in the interior of the Fermi/eROSITA bubbles, which themselves were formed by older activities in the Galactic center.
However, the HVMCs should also originate from a similar activity, which implies there are three independent activities, respectively correlated with the radio bubbles/X-ray chimneys, the HVMCs and the Fermi bubbles/eROSITA bubbles.
The difference is that the HVMCs will be difficult to propagate to much higher latitude in our simulations, because the acceleration rate of HVMCs at high latitude will largely decrease.
If the three independent activities are not related with each other, we have to use three models to respectively explain the structures at three scales, which will lead to an inelegant physical pattern.

Alternatively, as suggested by \citet{2019Natur.567..347P}, the X-ray chimney/the radio bubbles may be a channel that transports energy from the Galactic center to the high-latitude region currently occupied by the Fermi bubbles, and the HVMCs are the manifestation of the transportation process, which is a more elegant unified model.
In fact, the HVCs can spread from $\sim$ 100 pc to $\sim$ 10 kpc \citep{2018ApJ...855...33D, 2020ApJ...888...51L, 2022MNRAS.513.3228L}, though most of the FB HVCs are located in the lower 2 kpc, which may be the clue connecting the feedback relics at different scale.
In this case, the channel should have existed for tens of Myr, so that star formation in the Galactic center can be sufficient to supply the total energy content of the Fermi bubbles, $\sim 10^{56}$ erg \citep{2013Natur.493...66C}.
However, such a picture contradicts with the capped morphology of the radio bubbles (the southern bubble is not obviously capped in X-rays; \citealp{2021A&A...646A..66P}), which, according to our simulations, is naturally explained as the expanding shell of a newly born outflow.
This picture may be reconciled if star formation in the Galactic center has been episodic on a timescale of $\sim$10 Myrs \citep{2015MNRAS.453..739K}, then the X-ray chimney/the radio bubbles are (re)established and the HVCs/HVMCs are (re)accelerated by consecutive generations of mini-starbursts and collapses inbetween.
Of course, over such a long interval, the activity of Sgr A* can also play an important role in contributing to the formation of these relics, especially in view of the fact it was likely much more active in the recent past \citep{2010ApJ...714..732P, 2013ASSP...34..331P, 2018ApJ...856..180C}.
In a hybrid scenario, Sgr A*, with supernovae and even stellar winds, can simultaneously sustain the channel and transport energy to larger scales, implying X-ray emission beyond the edge of the radio bubbles, which is also suggested by \citet{2021A&A...646A..66P}.
For example, a AGN activity produces the large-scale structure and triggers the surrounding starburst, then the newly-formed massive stars drive strong stellar wind and explode as supernovae to produce the small-scale structure.
Possibly, the stellar winds and shock wave of supernovae can also trigger the tidal disruption event of the central SMBH, then produce a smaller-scale structure.

In conclusion, our findings suggest the existence of a potentially stable channel in the Galactic center, driven by a combination of diverse activities, which episodically accelerates gas clouds and transports energy to higher latitudes.
The HVMCs/FB HVCs are also the ingredient of the channel, but the HVMCs usually exist in low latitude due to the higher possibility of crushing at higher latitude. This pattern offers a comprehensive explanation for the interrelation between various feedback remnants, without necessitating the introduction of new models.

\section{Summary}\label{sec:sum}
To investigate the formation of HVMCs in our Galactic center, we perform simulations utilizing a starburst model, where HVMCs originate from low-latitude molecular clouds accelerated by a subsequent supernovae explosions. Previous studies have raised concerns about the destruction of molecular clouds due to the violent activity in the Galactic center, making it challenging for them to reach higher latitudes and velocities without disruption. However, our simulation results demonstrate that this problem can be resolved within a wide parameter space, given the appropriate local environment.

The main findings are summarized as follows:
\\
\begin{itemize}
\item{The HVMCs can indeed be formed in a starburst in the Galactic center.}

\item{The magnetic field can protect the molecular clouds.}

\item{The magnetic pressure, enhanced by the compression of shock wave, can contribute to accelerating the clouds.}

\item{The acceleration rate of HVMCs will largely decrease at high latitude, because the reverse shock, generated by the interaction between the shock wave and the molecular clouds,  can gradually balance the forward shock from the supernovae. Therefore, we can predict the largest velocity the HVMCs can reach is $\sim$ 400 km s$^{-1}$.}

\item{The mixture between the clouds and the ejecta of the supernovae is more efficient at low latitude, and this process can significantly impact the metallicity of HVCs.}

\item{HVMCs/FB HVCs potentially serve as ingredients in a channel sustained by diverse activities in the Galactic center, intermittently accelerating gas clouds and transporting energy to higher latitudes. }
\end{itemize}

Due to the limited size of the simulation box, the subsequent evolution of HVMCs beyond 1 kpc latitude remains uncertain. Furthermore, the small box size restricts us to initializing only one cloud, resulting in inconsistent HVC number density and distribution compared to observations. Future efforts involve expanding the simulation box, simplifying supernova explosion settings, and implementing adaptive mesh refinement to provide a more comprehensive understanding of the phenomenon.

\section*{Acknowledgements}
We acknowledge the cosmology simulation database (CSD) in the National Basic Science Data Center (NBSDC) and its funds the NBSDC-DB-10. We acknowledge the support from the National Key Research and Development Program of China (2022YFA1602903), from the National Science Foundation of China (12147103, 12273010), and from the Fundamental Research Funds for the Central Universities(226-2022-00216).

\section*{Data Availability}
The simulation data underlying this article may be shared upon reasonable request to the corresponding author.

\bibliographystyle{mnras}
\bibliography{mydb}

\appendix

\section{Cooling function}
\begin{figure}
\includegraphics[width=0.5\textwidth,keepaspectratio]{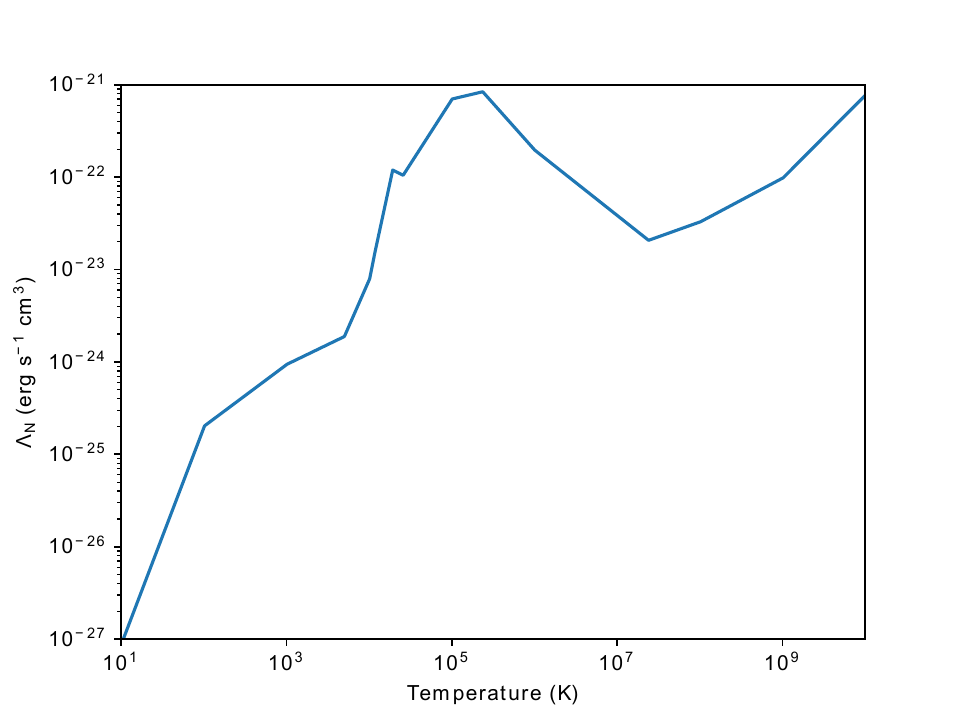}
\caption{The piece-wise cooling curve used in the simulations.
\label{fig:cool}}
\end{figure}

The cooling process can significantly influence the evolution of HVMCs, but an accurate tabulated cooling function will spend much more computational resource.
Therefore, we in the simulations adopt a piece-wise cooling function (see Figure~\ref{fig:cool}), which can roughly describe the cooling function.

\end{document}